\documentclass[12pt,a4paper]{article}
\usepackage{amsmath,amsthm}
\usepackage{amsmath,amssymb}

\numberwithin{equation}{section}

\newcommand {\Eg} {{\cal E}} 
\newcommand {\Fg} {{\cal F}}
\newcommand {\Hg} {{\cal H}} 
\newcommand {\Mg} {{\cal M}} 
\newcommand {\Ng} {{\cal N}} 
\newcommand {\Sg} {{\cal S}} 

\newcommand {\R} {{\mathbb R}}

\newcommand {\C} {{\mathbb C}} 

\newcommand {\Cb} {{\overline{\C}}}
\newcommand {\Bb} {{\bf B}}

\newcommand {\Ga} {\Gamma}
\newcommand {\p} {\psi}
\newcommand {\f} {\phi}
\newcommand {\ep} {\varepsilon}

\newcommand {\Th} {\Theta}

\newcommand {\ka} {\kappa}
\newcommand {\lam} {\lambda}
\newcommand {\m} {\mu}
\newcommand {\n} {\nu}
\newcommand {\s} {\sigma}
\newcommand {\Si} {\Sigma}
\newcommand {\w} {\omega}

\def\lbeq(#1){\label{eqn:#1}}
\def\refeq(#1){{\rm (\ref{eqn:#1})}}
\def\lbth(#1){\label{th:#1}}
\def\refth(#1){{\rm Theorem \ref{th:#1}}}
\def\refthb(#1){{\bf Theorem \ref{th:#1}}}
\def\refths(#1,#2){{\rm Theorems \ref{th:#1} and \ref{th:#2}}}
\def\lblm(#1){\label{lm:#1}}
\def\reflm(#1){{\rm Lemma \ref{lm:#1}}}
\def\reflms(#1,#2){{\rm Lemmas \ref{lm:#1} and \ref{lm:#2}}}
\def\reflmb(#1){{\bf Lemma \ref{lm:#1}}}
\def\lbprp(#1){\label{prp:#1}}
\def\refprp(#1){{\rm Proposition \ref{prp:#1}}}
\def\lbcor(#1){\label{cor:#1}}
\def\refcor(#1){{\rm Corollary \ref{cor:#1}}}

\newcommand {\bgth} {\begin{theorem}}
\newcommand {\edth} {\end{theorem}}
\newcommand {\bgprp} {\begin{proposition}}
\newcommand {\edprp} {\end{proposition}}
\newcommand {\bgdf} {\begin{definition}}
\newcommand {\eddf} {\end{definition}}
\newcommand {\bglm} {\begin{lemma}}
\newcommand {\edlm} {\end{lemma}}
\newcommand {\bgcor} {\begin{corollary}}
\newcommand {\edcor} {\end{corollary}}
\newcommand {\bgpf} {\begin{proof}}
\newcommand {\edpf} {\end{proof}}
\newcommand {\bgrm} {\begin{remark}}
\newcommand {\edrm} {\end{remark}}

\newcommand {\bqn} {\begin{equation}}
\newcommand {\eqn} {\end{equation}}

\newcommand {\ben} {\begin{enumerate}}
\newcommand {\een} {\end{enumerate}}
\newcommand {\ph} {\varphi}

\newcommand {\la} {\langle}
\newcommand {\ra} {\rangle}
\newcommand {\ds} {\displaystyle}
\newcommand {\ax} {{\la x \ra}}

\newcommand {\br} {\begin{array}}
\newcommand {\er} {\end{array}}
\newcommand {\lap} {\Delta}
\newcommand {\pa} {\partial}

\newtheorem{theorem}{Theorem}[section]
\newtheorem{lemma}[theorem]{Lemma}
\newtheorem{proposition}[theorem]{Proposition}
\newtheorem{definition}[theorem]{Definition}
\newtheorem{corollary}[theorem]{Corollary}

\theoremstyle{definition}
\newtheorem{remark}[theorem]{Remark}

\newcommand {\absleq} {{\leq_{|\, \cdot\, |}\, }}

\title{Wave Operators for Schr\"odinger Operators 
with Threshold Singuralities, Revisited}
\author{
K. Yajima\thanks{
Department of Mathematics, Gakushuin University, 
1-5-1 Mejiro, Toshima-ku, Tokyo 171-8588, Japan. 
Supported by JSPS grant in aid for scientific research 
No. 22340029 }}
\date{}
\begin{document}
\allowdisplaybreaks
\maketitle

\begin{abstract}
The continuity property in the Sobolev space 
$W^{k,p}(\R^m)$ of wave operators of 
scattering theory for $m$-dimensional single-body 
Schr\"odinger operator is considered when the resolvent 
of the operator has singularities at the bottom of the 
continuous spectrum. It is shown that they are continuous 
in $W^{k,p}(\R^m)$, $0\leq k \leq 2$ for $1<p<3$ but not 
for $p>3$ if $m=3$ and, 
for $1<p<m/2$ but not for $p>m/2$ if $m\geq 5$.  
This extends downward the previously known interval 
of $p$ for the continuity, $3/2<p<3$ for $m=3$ and 
$m/(m-2)<p<m/2$ for $m\geq 5$. The formula which 
represents the integral 
kernel of the resolvent of the even dimensional free 
Sch\"odinger operator as the superposition of 
exponential-polynomial like functions substantially 
simplifies the proof of the previous paper when 
$m \geq 6$ is even. 
\end{abstract}

\section{Introduction} 

Let $H_0=-\lap$ be the free Schr\"odinger operator
with natural domain $D(H_0)=H^2(\R^m)$ and 
$H= H_0 + V(x)$ with real $V(x)$ such that    
\begin{equation}\lbeq(decay-con)
|V(x)|\leq C\ax^{-\delta} \ \ \text{for some }\ \delta>2, \ 
\ax=(1+|x|^2)^\frac12. 
\end{equation}
We write $\Hg=L^2(\R^m)$ and 
$H^s(\R^m)=\{u \in \Hg \colon \pa^\alpha u \in \Hg, 
\ |\alpha|\leq s\}$ is the Sobolev space, $s\geq 0$ being 
an integer. Then, $H$ is selfadjoint in $\Hg$ with a core 
$C_0^\infty(\R^m)$ and it satisfies the following properties 
(see e.g. \cite{KatoS1,Ku-0,RSi,RS3,RS4}):  
\ben 
\item[{\rm (i)}]  The spectrum $\s(H)$ of $H$ consists 
of the absolutely continuous (AC for short) part 
$[0,\infty)$ and a finite number of non-positive 
eigenvalues $\{\lam_j\}$ of finite multiplicities. 
The singular continuous spectrum 
and positive eigenvalues are absent from $H$. 
\een
We write $\Hg_{ac}(H)$ for the AC spectral subspace 
of $\Hg$ for $H$ and $P_{ac}(H)$ is the orthogonal 
projections onto $\Hg_{ac}(H)$.   
\ben 
\item[{\rm (ii)}] Wave operators $W_\pm$ defined by 
strong limits 
\begin{equation}\lbeq(defw)
W_\pm u = \lim_{t \to\pm\infty} e^{itH}e^{-itH_0}u, 
\quad u \in \Hg  
\end{equation}
exist and are complete: ${\rm Image}\, W_\pm = \Hg_{ac}(H)$. 
They are unitary 
from $\Hg$ onto $\Hg_{ac}(H)$ and intertwine $H_{ac}$ 
and $H_0$: For Borel functions $f$,    
\begin{equation}\lbeq(inter)
f(H)P_{ac}(H) = W_\pm f(H_0) W_\pm^\ast.
\end{equation}
\een
If follows that various mapping properties of $f(H)P_{ac}$ 
may be deduced from those of $f(H_0)$ if the corresponding 
ones of $W_\pm$ are known. In particular, if 
\bqn \lbeq(lpbdd)
W_\pm \in \Bb(L^p(\R^m)) \quad \mbox{for} 
\  p \in [ p_1, p_2 ], 
\eqn 
then $W_\pm^\ast \in \Bb(L^q(\R^m))$ for 
$q\in [q_2,q_1]$, $1/p_j+1/q_j=1$, 
$j=1,2$, and   
\bqn \lbeq(pq)
\|f(H)P_{ac}(H)\|_{\Bb(L^q, L^p)} 
\leq C_{pq}\|f(H_0)\|_{\Bb(L^q, L^p)}, 
\eqn 
for these $p$ and $q$ with a constant $C_{pq}$ which 
is {\it independent of $f$}. 
Here $f(H_0)$ is a Fourier multiplier by 
$f(\xi^2)$ and its mapping properties should be much 
easier to investigate than those of $f(H)P_{ac}(H)$. 
We define the Fourier transform $\hat{u}(\xi)=\Fg{u}(\xi)$ 
and the conjugate one $\Fg^\ast{u}(\xi)$ by    
\[
\Fg u(\xi) = \int_{\R^m} e^{-ix\xi} u(x) dx\ \mbox{and} \  
\Fg^\ast u(\xi) = \frac1{(2\pi)^m}
\int_{\R^m} e^{ix\xi} u(x) dx
\] 
respectively. The intertwining property \refeq(inter) 
may be made more precise: Wave operators $W_\pm$ are 
transplantations (\cite{Stempak}) of the complete set  
of (generalized) eigenfunctions 
$\{e^{ix\xi} \colon \xi \in \R^m\}$ of $-\lap$ by 
those of out-going and in-coming scattering eigenfunctions 
$\{\ph_{\pm}(x,\xi) \colon \xi\in \R^m\}$ of 
$H=-\lap + V$ (\cite{Ku-0}):  
\[
W_{\pm}u(x) = \Fg_{\pm}^\ast \Fg u(x)=
\frac1{(2\pi)^d}\int_{\R^d} \ph_{\pm}(x,\xi) 
\hat u(\xi) d\xi, 
\]
where $\Fg_{\pm}$ and $\Fg_\pm^\ast$ are the generalized 
Fourier transforms associated with 
$\{\ph_{\pm}(x,\xi) \colon \xi\in \R^m\}$ 
and the conjugate ones respectively defined by 
\[
\Fg_{\pm} u(\xi) = \int_{\R^d}\overline{\ph_{\pm}(x,\xi)} 
u(x) dx, \quad 
\Fg_{\pm}^\ast u(\xi) = 
\frac1{(2\pi)^m} 
\int_{\R^d}\ph_{\pm}(x,\xi)u(x) dx.
\]
They satisfy the inversion formulas: 
$\Fg_{\pm}^\ast \Fg_{\pm} u = u$ for $u \in \Hg(H)$ .
It follows for multiplication operators $M_F$ with 
Borel functions $F(\xi)$ on $L^2(\R^m)$ that 
\[
F(D_\pm) \equiv \Fg_{\pm}^\ast M_F \Fg_{\pm}u = 
W_\pm F(D)W_{\pm}^\ast u , \quad u \in \Hg_{ac}(H) 
\]
and the estimate \refeq(pq) remains valid when  
$f(H)$ and $f(H_0)$ are replaced by $F(D_\pm)$ and 
$F(D)$. 

In this paper we are interested in the problem that 
under what conditions $W_\pm$ are bounded in $L^p(\R^m)$, 
which almost automatically implies that they are bounded 
also in $W^{k,p}(\R^m)$, $\leq k\leq 2$ (see below). 

There is now a substantial literature on this problem 
(\cite{Be, DF,FY, JY, Weder, Y-d3, Y-odd}) and it is 
known that the answer depends on the spectral 
properties of $H$ at the bottom of the AC spectrum of $H$. 
We define 
\bqn 
\Eg=\{u \in H^2(\R^m) \colon (-\lap + V)u = 0\}, 
\eqn 
the eigenspace of $H$ associated with the eigenvalue $0$ and 
\begin{equation}\lbeq(thres)
\Ng= \{u \in \ax^{s}L^2(\R^m): 
(1 + (-\lap)^{-1}V) u=0\} = 0 . 
\end{equation} 
The space $\Ng$ is independent of $1/2<s<\delta-1/2$ 
and $\Eg \subset \Ng$ (\cite{JK}). The operator $H$ is 
said be of {\it generic type} if 
$\Ng=\{0\}$ and of {\it exceptional type} otherwise. 
When $H$ is of {\it generic type}, we have rather 
satisfactory results (though there is much space for 
improvement in the condition on $V$) and it 
has been proved that $W_\pm$ are bounded in $L^p(\R^m)$ 
for all $1\leq  p \leq \infty$ if $m \geq 3$ and, for all 
$1<p<\infty$ if $m=1$ and $m=2$ under various smoothness 
and decay at infinity assumptions on $V$ (see \cite{Be} 
for the best result when $m=3$); but they are 
in general {\it not} bounded in $L^1(\R^1)$ or 
$L^\infty(\R^1)$ when $m=1$ (\cite{Weder}). 
When $H$ is of exceptional type, it is long known that the 
same results hold when $m=1$ (see \cite{Weder, AY, DF}), 
however, we still poorly understand the problem 
when $m \geq 2$. It is proved that $W_\pm$ are bounded 
in $L^p(\R^m)$ for $p\in (m/(m-2), m/2)$ when $m \geq 5$ 
and for $3/2<p<3$ when $m=3$ (\cite{FY,Y-odd}, see also 
\cite{JY} for a partial result for $m=4$) 
but nothing more to the best knowledge of the author.  
The purpose of this paper is to improve this 
situation by proving that $W_\pm$ are actually bounded 
in $L^p(\R^m)$ for a wider range of $p$'s when 
$m=3$ and $m\geq 5$. The cases $m=2, 4$ and the end point 
problem are left for future investigation. 

Throughout the paper, we assume that $V$ is real 
measurable function which satisfies 
the following conditions, where $m_\ast=(m-1)/(m-2)$ and 
$C>0$ and $\ep>0$ are constants: 
\begin{gather} \lbeq(cond-1)
\Fg (\ax^{2\s} V) \in L^{m_\ast}  \quad 
\mbox{for some}\  \s>1/m_\ast.  \\ 
|V(x)|\leq C \ax^{-\gamma}, \quad 
\gamma= \left\{\br{ll} m+4+\ep, 
\quad &  \mbox{if}\ \ 3\leq m\leq 7, \\ 
m+3+\ep \quad & \mbox{if}\ \ m\geq 8. \er \right. 
\lbeq(decay) 
\end{gather} 
Note that the condition \refeq(cond-1) requires certain 
smoothness for $V$. 
 
\begin{theorem}\lbth(theo) 
Let $m\geq 3$. Suppose that $V$ satisfy \refeq(cond-1) 
and \refeq(decay) and that $H$ is of exceptional type. 
Then: 
\ben 
\item[{\rm (1)}] 
For any $0\leq k \leq 2$ and $p$ such that  
$\left\{ \br{ll} 1<p<3, \ & \mbox{if}\ m=3, \\
1<p<m/2,  & \mbox{if}\ m\geq 5 \er \right. $
\begin{equation}\lbeq(bounded-1) 
\|W_\pm u \|_{W^{k,p}} 
\leq C_p \|u\|_{W^{k, p}}, \quad 
\ u \in W^{k,p}(\R^m) \cap L^{2}(\R^m) 
\end{equation} 
\item[{\rm (2)}]  
If $m=3$ and $p>3$ or if $m\geq 5$ and $p>m/2$, then 
$W_\pm \not\in \Bb(L^{p}(\R^m))$.  
\een
\end{theorem}

We should emphasize that we have no results 
for $p$ at the end points of the intervals and 
that the problem is completely open when $m=2, 4$. 

The rest of the paper is devoted to the proof of 
\refth(theo). We shall be concentrated in the proof 
of statement (1) because the second statement is 
contained in the result of Murata (\cite{Mu}, Theorem 1.2) 
which implies in particular that there exists a 
$u \in C_0^\infty(\R^m)$ such that 
\bqn 
\lim_{t\to \infty} 
t^{m\left(\frac12-\frac1{p}\right)}  
\|e^{-itH}P_{ac}(H) u \|_{p}= \infty   
\eqn 
for any $p>3$ if $m=3$ and for any $p\in (m/2, \infty)$ if 
$m \geq 5$, which contradicts with \refeq(bounded-1) 
which would imply 
\bqn   
\|e^{-itH}P_{ac}(H) u \|_{p} \leq C 
t^{m\left(\frac12-\frac1{p}\right)} \|u\|_{p'}, 
\eqn 
for the dual exponent $p'$ of $p$ by virtue of the well 
known $L^p$-$L^q$ estimate for the free propagator 
$e^{-itH_0}$ and the intertwing property \refeq(inter). 
We also remark that statement (1) of the theorem 
follows if we prove 
\bqn \lbeq(bound-2)
\|W_\pm u \|_{L^p} \leq C_p \|u\|_{L^p}, \quad u \in 
C_0^\infty(\R^m).
\eqn 
In fact the norms 
$\|(H_0+c^2)u \|_{L^p}$ and $\|u\|_{W^{2,p}}$ are equivalent 
for any $1<p<\infty$ and $c>0$ via the Riesz transform and 
$\|(H+c^2)u\|_p $ and $\|(H_0+c^2)u\|_p $ are also 
equivalent for sufficiently large $c>0$ since we have 
$(H+V+c^2)= (1+ V(H_0+c^2)^{-1})(H_0+c^2)$ 
and $(1+ V(H_0+c^2)^{-1})$ is an isomorphism of $L^p(\R^m)$. 
Thus, the intertwining property 
and \refeq(bound-2) imply 
\begin{multline} \lbeq(bound-3)
\|W_\pm u \|_{W^{2, p}}
\leq C \|(H+i)W_\pm u \|_{L^p}  
=\|W_\pm (H_0+i)u \|_{L^p} \\ 
 \leq C_p \|(H_0+i)u\|_{L^p} \leq 
C \|u\|_{W^{2,p}}, \quad C_0^\infty(\R^m), 
\end{multline}
and the estimate \refeq(bounded-1) for $k=1$ then follows 
by interpolation. Thus, we shall be devoted 
in what follows to proving \refeq(bound-2), the 
$L^p$-boundedness of $W_\pm$. 

We use the following notation and conventions: 
\[
f \, \absleq \, g \ \mbox{means}\ |f|\leq |g|.
\]
For functions $u$ and $v$, 
whenever $\overline{u(x)}v(x)$ is integrable we write 
\[
\la u, v \ra = \int_{\R^n} \overline{u(x)}v(x) dx.
\]
We use this notation when $v$ is in a certain function space 
and $u$ is in its dual space. 
The rank $1$ operator $\f \mapsto \la v, \f \ra u $ 
is interchangeably denoted by
\[
|u\ra \la v| = u \otimes v .
\]
For Banach spaces $X$ and $Y$, $\Bb(X,Y)$ 
is the Banach space of bounded operators from $X$ to $Y$ 
and $\Bb(X)=\Bb(X,X)$. $\Bb_\infty(X,Y)$ and 
$\Bb_\infty(X)$ are spaces of 
compact operators. The identity operators in various 
Banach spaces are indistinguishably denoted by $1$. 
For $1\leq p \leq \infty$, 
$\|u\|_p=\|u\|_{L^p}$ is the norm of the Lebesgue 
spaces $L^p(\R^m)$. When $p=2$, we often omit $p$ 
and write $\|u\|$ for $\|u\|_2$. 
For $s \in \R$, 
\[
L^2_{s}=\ax^{-s}L^2=
L^2(\R^m, \ax^{2s}dx), \quad H^s(\R^m)=\Fg L^2_s(\R^m)
\]
are the weighted $L^2$ spaces  and Sobolev spaces. 
The space of rapidly decreasing functions is 
denoted by $\Sg(\R^m)$. 

We denote the resolvents of $H$ and $H_0$ respectively by  
\[
R(z)= (H-z)^{-1}, \quad R_0(z) = (H_0 -z)^{-1}.
\]
We parameterize $z \in \C \setminus [0,\infty)$ 
as $z=\lam^2$ by $\lam \in \C^+=\{z \in \C: \Im z >0 \}$ 
so that the boundaries $\{\lam \colon \pm \lam \in (0,\infty)\}$ 
are mapped onto the upper and lower edge of the positive 
half line $\{z\in \C \colon z>0\}$. We define 
\[
G(\lam) = R(\lam^2), \quad G_0(\lam) = R_0(\lam^2), 
\quad  \lam \in \C^+.
\]
These are $\Bb(\Hg)$-valued meromorphic functions of 
$\lam \in \C^+$ and the limiting absorption principle 
\cite{Ku-0} (LAP for short) asserts that, when considered as 
$\Bb(\ax^{-s}L^2, \ax^{-t}L^2)$-valued 
functions, $s, t>\frac12$ and $s+t>2$, $G_0(\lam)$ 
has the locally H\"older continuous extensions to its closure 
$\Cb^+=\{z: \Im z \geq 0\}$. The same is true also for  
$G(\lam)$, but, if $H$ is of exceptional type, 
it has singularities at $\lam=0$. In what follows $z^\frac12$ 
is the branch of square root of $z$ cut along the negative 
real axis such that $z^\frac12>0$ when $z>0$. We assume 
$m \geq 3$ in what follows. 

{\bf Acknowledgement} \ The part of this work was carried 
out while the author was visiting Aalborg university 
and Aarhus university. He would like to express his 
sincere thanks to both institutions for the hospitality, 
to Arne Jensen, Jacob Schach Moeller and Erik Skibsted 
in particular.

\section{Reduction to the low energy analysis}  
We study $W_{-}$ only and denote it by $W$ for simplicity. 
When $u \in \ax^{-s}L^2$, $s>1/2$, we may represent 
$Wu$ as the limit 
\begin{align} 
Wu & =u-\lim_{\ep \downarrow 0, N\uparrow \infty}
\frac{1}{\pi i}\int^N_{\ep}  
G(\lam)V(G_0(\lam)-G_0(-\lam))u \lam d\lam
\lbeq(stationary-0) \\
& = u -\frac{1}{\pi i}\int^\infty_{0}  
G(\lam)V(G_0(\lam)-G_0(-\lam))u\lam d\lam 
\lbeq(stationary)
\end{align} 
by using the boundary values of the resolvents 
(\cite{Ku-0}). Here we understand the integral on the 
right of \refeq(stationary-0) as the Riemann integral 
of an $\ax^{t}L^2$-valued continuous function, 
where $t>1/2$ is such that $s+t>2$. Then, the 
result of integral belongs to $L^2(\R^m)$, the 
limit exists in $L^2(\R^m)$ and the equation 
\refeq(stationary-0) is satisfied, which we symbolically 
write as \refeq(stationary). 

We decompose $W$ into the high and the low energy parts 
\bqn \lbeq(lh-decom)
W=W_{>} + W_{<} \equiv W\Psi(H_0) + W \Phi(H_0),
\eqn 
by using cut off functions $\Phi\in C_0^\infty(\R)$ 
and $\Psi\in C^\infty(\R)$ such that 
\[
\Phi(\lam^2) + \Psi(\lam^2) \equiv 1, 
\quad \Phi(\lam^2)=1 \ \mbox{near $\lam=0$  and 
$\Phi(\lam^2)=0$ for $|\lam|>\lam_0$}
\] 
for a small constant $\lam_0>0$  and we study $W_{<}$ and 
$W_{>}$ separately. 
We have proven in previous papers \cite{Y-odd, FY} 
that, under the assumption of this paper, the high energy 
part $W_{>}$ is bounded in $L^p(\R^m)$ for all 
$1\leq p \leq \infty$ if $m \geq 3$ and we have nothing to add 
in this paper for $W_{>}$. Thus, in what follows, we shall 
be devoted to studying the low energy part $W_{<}$ 
\bqn 
W_{<} = \Phi(H_0)u  -
\frac{1}{\pi i}\int^\infty_{0}  
G(\lam)V(G_0(\lam)-G_0(-\lam))\lam \Phi(H_0) d\lam. 
\lbeq(stationary-l)
\eqn 
Since $\Phi(H_0)$ is evidently bounded in $L^p(\R^m)$ for all 
$1\leq p \leq \infty$, 
we have only to study the integral part of 
\refeq(stationary-l) which we may write in the form 
\bqn \lbeq(z)
Zu = -\frac{1}{\pi i}\int^\infty_{0}  
G_0(\lam)V(1+ G_0(\lam)V)^{-1}
(G_0(\lam)-G_0(-\lam))\lam F(\lam)u d\lam 
\eqn 
by using \reflm(mulfo) below. where $F(\lam)= \Phi(\lam^2)$. 
When $H$ is of generic type, we have shown also in 
\cite{Y-odd, FY} that $Z$ is bounded in $L^p(\R^m)$ 
for all $1\leq p \leq \infty$ under the condition of 
\refth(theo). Therefore, {\it we assume in the rest of the paper 
that $H$ is of exceptional type}.  

\subsection{Low energy asymptotics. Further reduction}

Since $\delta>2$, the LAP (cf. Lemma 2.2 of \cite{Y-odd}) 
implies that $G_0(\lam)V$ is a locally H\"older continuous 
function of $\lam \in \R$ with values in $\Bb_\infty (L^{-s})$ 
for any $1/2<s<\delta-1/2$ and, the absence 
of positive eigenvalues (\cite{Kato-e}) implies 
that $1+ G_0(\lam)V$ is 
invertible for $\lam>0$ (cf. \cite{Ag}). It follows from the 
resolvent equation $G(\lam)= G_0(\lam)-G_0(\lam)V G(\lam)$ that   
\bqn 
G(\lam)V= G_0(\lam)V(1+ G_0(\lam)V)^{-1}\ \mbox{for} \ \lam \not=0  
\eqn 
and, it also is $\Bb_\infty (L^{-s})$-valued locally H\"older 
continuous for $\lam\in \R \setminus\{0\}$. However, if $H$ 
is of exceptional type 
$\Ng= {\rm Ker}_{L^{-s}} (1 + G_0(0)V)\not=0$, and 
$(1+ G_0(\lam)V)^{-1}$ has singularities at $\lam=0$. 
We determine their singularities by recalling 
some results from \cite{Y-odd} and \cite{FY} and further 
reduce the problem to the study of $Z_s$ which is obtained 
by inserting the singular part of $(1+ G_0(\lam)V)^{-1}$ into 
\refeq(z). 

Since $G_0(0)V \in \Bb_{\infty}(L^2_{-s})$ 
with real integral kernel $C_m|x-y|^{2-m}V(y)$, $C_m>0$,  
$\Ng$ is of finite dimensional and we may choose 
real valued functions as a basis. Functions $u \in \Ng$  
satisfy the equation 
\bqn 
-\lap u + V u =0, 
\eqn 
hence $u \in \ax^{-s} H^2(\R^m)$ for any $s>1/2$ and, 
moreover,  
\bqn \lbeq(As)
|u(x)\leq C \ax^{2-m}, \ \mbox{hence, in particular}\ \  
Vu \in L^1(\R^n)\cap L^\infty(\R^m)
\eqn  
(see e.g. \cite{AS}, corollary 2.6), 
however, it may 
not be an eigenfunction if $m\leq 4$. 

\subsubsection{Odd dimensional cases}
The structure of singularities are different for different 
$m$. For odd dimensions $m \geq 3$ we have the following 
results (see, e.g. Theorem 2.12 of \cite{Y-odd}). We state it 
separately for $m=3$ and $m \geq 5$.  

\paragraph{The case $m =3$} If $m=3$, $u \in \Ng$ satisfies 
\[
u(x) = -\frac1{4\pi}\int_{\R^3} \frac{V(y)u(y)}{|x-y|}dy. 
\]
It follows that    
\bqn \lbeq(asym-3)
u(x) = \frac{a(u)}{|x|} + O(|x|^{-2}) \ 
\mbox{as $|x|\to \infty$}, \quad 
a(u) = \frac1{4\pi}\int_{\R^3}V(x)u(x) dx.
\eqn 
Thus, 
$\Eg= \{u \in \Ng\setminus\{0\} \colon a(u)=0\}$ 
and,  as $\Ng \ni u \mapsto a(u)\in \C$ 
is a continuous functional, $\dim \Ng/\Eg \leq 1$. Any 
$\ph \in \Ng \setminus \Eg$ 
is called {\it threshold resonance} of $H$. 
We say that $H$ is of exceptional type 
of {\it the first kind} if $\Eg=\{0\}$, 
{\it the second} if $\Eg=\Ng$ and {\it the third kind} if 
$\{0\}\subsetneq \Eg \subsetneq \Ng$. The orthogonal projection 
in $\Hg$ 
onto the eigenspace $\Eg$ will be denoted by $P_0$.
We let $D_0, D_1, \dots$ be the integral operators defined by 
\[
D_j u(x) = \frac1{4{\pi}j!} \int_{\R^3}|x-y|^{j-1}u(y) dy, 
\quad j=0,1, \dots.
\]
so that we have a formal Taylor expansion  
\[
G_0(\lam)u(x)= \frac1{4\pi}\int_{\R^3}
\frac{e^{i\lam|x-y|}}{|x-y|}u(y)dy 
=\sum_{j=1}^\infty (i\lam)^j D_j u .
\]
If $H$ is of exceptional type of the third kind, there 
a unique $\p\in \Ng$ such that 
\[
-(V\p, u)=0, \quad \forall u \in \Eg, \quad 
-(V\p,\p)=1 \ \mbox{and}\ a(\p)>0. 
\]
We define 
\[
\ph=\p + P_0 VD_2 V \p\in \Ng
\] 
and call it {\it the canonical resonance}. 
If $H$ is of exceptional type of the first kind, 
then $\dim\Ng=1$ and there is a 
unique $\ph\in \Ng$ such that $-(V\ph, \ph)=1$ 
and $a(\ph)>0$ and we call this the canonical resonance. 
We have the following result for $m=3$ 
(see e.g. \cite{Y-odd}). $\Bb_2(\Hg)$ is 
the Hilbert space of Hilbert-Schmidt operators on $\Hg$. 

\begin{theorem}\lbth(III) Let $m=3$ 
and let $V$ satisfy $|V(x)|\leq C\ax^{-\delta}$ 
for some $\delta>3$. If $H$ is of exceptional 
type of the third kind, $\ph$ is the canonical 
resonance and $a = 4\pi i |\la V,\ph \ra|^{-2}$,   
Then, for any $s\in (3/2,\delta-3/2)$, 
\begin{gather}\lbeq(res)
(I+G_0(\lam)V)^{-1}
=\frac{P_0V}{\lam^2} - \frac{P_0V D_3VP_0V}{\lam}  
- \frac{a}{\lam}|\varphi\ra \la \varphi| V  + I + L(\lam). \\
\lam \mapsto 
\ax^{-s} L(\lam)\ax^{\delta-s}\in \Bb_2(\Hg) \ 
\mbox{is of class $C^{\s-\frac32}$ for any $\s<s$.} \lbeq(res-a)
\end{gather}
If $H$ is of exceptional type of 
the first or the second kind, \refeq(res) and \refeq(res-a) 
hold with $P_0=0$ or $\ph=0$ respectively. 
\end{theorem}

\paragraph{The case $m \geq 5$} 
If $m\geq 5$, \refeq(As) implies there are no threshold 
resonances and $\Ng=\Eg$. We write $P_0$ for the 
orthogonal projection in $\Hg$ onto $\Eg$. 
  
\begin{theorem}\lbth(m=5) Let $m\geq 5$ be odd and  
$|V(x)|\leq C\ax^{-\delta}$ for some $\delta>m+3$.  
Suppose $H$ is of exceptional type. Then, there exists 
an $L(\lam)$ which satisfies the property \refeq(res-a) 
for any $s\in (3/2,\delta-3/2)$ with $\Bb_\infty(\Hg)$ 
replacing $\Bb_2(\Hg)$ such that, if $m=5$,  
\bqn 
(I+G_0(\lam)V)^{-1}
=\frac{P_0V}{\lam^2} - 
\frac{c_0}{\lam}|\varphi\ra \la \varphi| V  
+ I + L(\lam) ,
\lbeq(res-5)
\eqn 
with $\ph =P_0 V$, $V$ being 
considered as a function, and $c_0=i/(24\pi^2)$;  
and if $m\geq 7$, 
\bqn 
(I+G_0(\lam)V)^{-1} = \frac{P_0V}{\lam^2} + I + L(\lam). 
\lbeq(res-7)
\eqn  
\end{theorem}

\paragraph{Decomposition of $Z$ for odd dimensions.} 
Write the right of \refeq(res), \refeq(res-5) and 
\refeq(res-7) as $S(\lam) + I + L(\lam)$ and insert 
this for $(I+G_0(\lam)V)^{-1}$ in the right side of 
\refeq(z). This yields $Zu= Z_{r1}u + Z_{r2}u + Z_{s} u$ 
where 
\begin{align}
Z_{r1}& = \frac{i}{\pi}\int^\infty_0 
G_0(\lam)V(G_0(\lam)-G_0(-\lam))F(\lam)\lam d\lam, \lbeq(wr0) \\
Z_{r2}& = \frac{i}{\pi}\int^\infty_0 
G_0(\lam)VL(\lam)(G_0(\lam)-G_0(-\lam))F(\lam)\lam d\lam,\lbeq(wr) \\
Z_{s}& = \frac{i}{\pi}\int^\infty_0 G_0(\lam)
VS(\lam)(G_0(\lam)-G_0(-\lam))F(\lam)\lam d\lam. \lbeq(zs) 
\end{align} 
We have shown that $Z_{r1}, Z_{r2}\in \Bb(L^p)$ 
for any $1\leq p \leq \infty$ 
in \cite{Y-d3} and in Sec. 3.1 of \cite{Y-odd} 
respectively. Thus, when $m \geq 3$ is odd, 
we have only to investigate the operator $Z_{s}$.  

\subsection{Even dimensional case} 
In even dimensions, singular terms may contain 
logarithmic factors. The following theorem is 
Proposition 3.6 of \cite{FY}. We let $\dim \Eg=d$. 

\bgth Let $m\geq 6$ be even. Suppose 
$|V(x)|\leq C \ax^{-\delta}$ for $\delta>m+4$ if $m=6$ 
and for $\delta>m+3$ if $m\geq 8$. Then, with an operator 
valued function $E(\lam)$ which produces a bounded operators 
in $L^p(\R^m)$ for any $1\leq p \leq \infty$ when inserted 
into \refeq(z) for $(1+ G_0(\lam)V)^{-1}$, we have the 
following statements: 
\ben 
\item[{\rm (1)}] If $m=6$, then we have  
\begin{equation}\lbeq(sing-1) 
(1+ G_0(\lam)V)^{-1}= \frac{P_0 V}{\lam^2} + 
\sum_{j=0, 1} \sum_{k=1,2} 
D_{jk} \lam^j \log^k \lam + E(\lam), 
\end{equation}
where all $D_{jk}$ are of rank at most $2d$ and $VD_{jk}$ 
are of the form 
\bqn \lbeq(sing-ef) 
VD_{jk}= \sum_{a,b=1}^{2d} \ph_a \otimes \p_b, 
\quad \ph_a,\ \p_b \in \ax^{-\delta+3+\ep}H^2(\R^6), \quad 
\forall \ep>0.  
\eqn 
\item[{\rm (2)}] If $m\geq 8$, with a constant $c_m$ 
and function $\ph = P_0 V$ with $V$ being considered 
as a function,     
\begin{equation}\lbeq(sing-2) 
(1+ G_0(\lam)V)^{-1}= 
\frac{P_0 V}{\lam^2} + 
c_m \ph \otimes (V\ph) \lam^{m-6} \log \lam + E(\lam). 
\end{equation}
If $m \geq 12$, then 
$c_m \ph \otimes (V\ph) \lam^{m-6} \log \lam$ may 
be included in $E(\lam)$. 
\een
\edth 
\bgrm In fact for $m=6$, we have proven in \cite{FY} only 
that $D_{jk}$ are of the form 
$D_{jk}=P_0 V D_{jk}^{(1)}P_0V +D_{jk}^{(2)}P_0V + 
P_0 V D_{jk}^{(3)}$ and, for any $\ep>0$, 
\bqn \lbeq(repla)
D_{jk}^{(1)}\in \Bb(\Ng), \  
D_{jk}^{(2)}\in\Bb(\Ng,\ax^{-3-\ep}L^2), \    
D_{jk}^{(3)}\in\Bb(\ax^{-\delta+3+\ep}L^2,\Ng).
\eqn 
However, $1+ G_0(\lam)V$ is a locally H\"older continuous 
function of $\lam\in \R$ with values in $\Bb(\ax^{-s}H^2(\R^6))$ 
and the proof of Proposition 3.6 goes through with 
$\ax^{-\gamma}H^2(\R^6)$ replacing $\Hg_{-\gamma}$ 
everywhere, which implies that \refeq(repla) holds with 
$H^2(\R^6)$ replacing $L^2$. This implies \refeq(sing-ef) 
by virtue of Riesz representation theorem. 
\edrm

By inserting \refeq(sing-1) 
and \refeq(sing-2) for $(I+G_0(\lam)V)^{-1}$ in the right 
of \refeq(z), we have the decomposition 
$Zu= Z_{r}u + Z_{\log}u + Z_{s} u$ 
as in the case $m$ is odd and    
\[
Z_{r} = \frac{i}{\pi}\int^\infty_0 
G_0(\lam)V E(\lam)(G_0(\lam)-G_0(-\lam))F(\lam)\lam d\lam 
\]
is bounded in $L^p(\R^m)$ for any $1\leq p \leq \infty$ 
(\cite{FY}). Thus, we need study 
\begin{align}
Z_{\log}& = \sum_{j,k} \frac{i}{\pi}\int^\infty_0 
G_0(\lam)V \lam^j (\log \lam)^k D_{jk}
(G_0(\lam)-G_0(-\lam))F(\lam)\lam d\lam,\lbeq(loge) \\
Z_{s}& = \frac{i}{\pi}\int^\infty_0 G_0(\lam)
VP_0V(G_0(\lam)-G_0(-\lam))F(\lam)\lam^{-1} d\lam \lbeq(e-zs) 
\end{align} 
in what follows for even $m\geq 6$.
 
\section{Preliminaries} 

We record several results which we need in what follows 
and which are mostly well known. 

\subsection{Results from harmonic analysis.}  

The following is the Muckenhaupt 
weighted inequality (cf. \cite{Gr}, Chapter 9). 

\begin{lemma}\lblm(ap) The weight $|r|^{a}$ is an  
$A_p$ weight on $\R$ if and only 
if $-1< a <p-1$. 
The Hilbert transform $\tilde \Hg$ and the 
Hardy-Littlewood maximal operator 
$\Mg$ are bounded in $L^p(\R, w(r)dr)$ for 
$A_p$ weights $w(r)$.  
\end{lemma} 
We shall repeatedly use the fact that   
\[
\mbox{$|r|^{m-1-p(m-1)}$,\ \ $|r|^{m-1-2p}$ \ and \ 
$|r|^{2-p}$}
\] 
are $A_p$ weights on $\R^1$ respectively 
for $1<p<m/(m-1)$ ($m\geq 2$), for 
$m/3<p<m/2$ ($m\geq 3$) and for $3/2<p<3$.  

For a function $F(x)$ on $\R^m$, $G(|x|)\in L^1(\R^m)$ 
is said to be a radial decreasing integrable majorant 
(RDIM) of $F$ if $G(r)>0$ is a decreasing function of $r>0$, 
and $|F(x)| \leq G(|x|)$ for a.e. $x\in \R^m$. The 
following lemma may be found e.g. on page 57 of \cite{Stein}. 

\bglm \lblm(max) 
\ben
\item[{\rm (1)}]
A rapidly decreasing function $F \in \Sg(\R^m)$ has a RDIM.
\item[{\rm (1)}] If $F$ has a RDIM. then there is a constant 
$C>0$ such that 
\begin{equation}\lbeq(max)
|(F \ast u)(t)| \leq C(\Mg u)(t), 
\quad t \in \R. 
\end{equation}
\een
\edlm 

We define the operator $\Hg$ on $\R$ by 
\bqn \lbeq(Hg) 
\Hg u(\rho ) = \frac{(1+\tilde \Hg)u(\rho )}{2} 
= \frac1{2\pi}\int_0^\infty e^{ir\rho } \hat{u}(r)dr 
\eqn 
where $\tilde{\Hg}$ is the Hilbert transform.

\bglm \lblm(add-cut) 
For $u$ and $F \in L^1(\R)$ such that 
$\hat{u}, \, \hat{F} \in L^1(\R)$ we have 
\bqn \lbeq(add-cut)
\frac1{2\pi}\int_0^\infty e^{i\lam\rho }
F(\lam) \hat{u}(\lam ) d\lam =  (\Fg^\ast{F}\ast \Hg u)(\rho ). 
\eqn 
\edlm 
\bgpf Let $\Th(\lam)=\left\{\br{l} 1, \ \mbox{for}\ \lam>0  \\ 
0, \ \mbox{for}\ \lam\leq 0 \er \right. $.   
Then, the left side equals  
\begin{multline*}
\frac1{2\pi}\int_{\R} e^{i\lam\rho }
F(\lam)\Th(\lam)\hat{u}(\lam) d\lam= 
\frac1{2\pi}\int_{\R}\left(
\int_{\R} e^{i\lam(\rho -\xi)}\Fg^\ast{F}(\xi)d\xi
\right) \Th(\lam)\hat{u}(\lam) d\lam\\
=\frac1{2\pi}\int_{\R} \Fg^\ast {F}(\xi) 
\Fg\{\Th(\lam)\hat{u}(\lam)\}(\rho -\xi) d\xi 
=(\Fg^\ast{F} \ast \Hg u)(\rho ) 
\end{multline*} 
as desired. 
\edpf

\subsection{Resolvent kernel}
The resolvent $G_0(\lam)$ for $\Im \lam \geq 0$ is 
a convolution operator and the convolution kernel is 
given for $m\geq 2$ by the Whittaker formula (cf. \cite{WW}) 
\begin{equation}\lbeq(co-ker) 
G_0(\lam,x) = \frac{e^{i\lam|x|}}{2(2\pi)^{\frac{m-1}{2}}
\Ga\left(\frac{m-1}{2}\right)|x|^{m-2}}
\int^\infty_0 e^{-t}t^{\frac{m-3}2}
\left(\frac{t}2-i\lam|x|\right)^{\frac{m-3}2}dt.  
\end{equation}
When $m\geq 3$ is odd, $\frac{m-3}2$ is an integer 
and we have the following expression. 
\bglm \lblm(odd-green) Let $m \geq 3$ be odd. Then,  
$G_0(\lam, x)$ is an exponential polynomial like function: 
\begin{equation}\lbeq(coker)
G_0(\lam,x) = \sum_{j=0}^{(m-3)/2} 
C_j \frac{(\lam|x|)^j e^{i\lam |x|}}{|x|^{m-2}} \ \mbox{with}\ 
C_j=\frac{(-i)^j(m-3-j)!}
{2^{m-1-j}\pi^{\frac{m-1}2}j! (\frac{m-3}2 -j)!}. 
\end{equation} 
and the coefficients $C_0$ and $C_1$ satisfy 
\bqn 
iC_0 +C_1=0, \quad \mbox{when $m \geq 5$}.
\eqn 
\edlm 

If $m$ is even, then $\frac{m-3}{2}$ is a half integer 
and derivatives of $G_0(\lam, x)$ become singular at 
$\lam=0$. This makes the analysis for even dimensional 
cases considerably more complex than for odd 
dimensional ones. Nevertheless, the expression given below 
of $G_0(\lam,x)$ for even dimensions $m\geq 4$ as a 
superposition of exponential polynomial like functions 
allows some arguments for even dimensions to go in parallel 
to the ones used for odd dimensional cases. We set 
\[
\n= \frac{m-2}2 
\]
and define superposing operators $T_j^{(a)}$ 
over parameter $a>0$ for $j=0, \dots, \n$ by 
\begin{gather} \lbeq(sum-j)
T_j^{(a)}[f(x,a)] = C_{m,j}
\int_{\R_{+}} (1+a)^{-(2\n-j+\frac12)} 
f(x,a) \frac{da}{\sqrt{a}}, \\
C_{m,j}= (-2i)^j\frac{\Ga\left(2\n-j+\frac12\right)}{
2^{m-1}\pi^{\frac{m}{2}}\Ga\left(\frac{m-1}{2}\right)}
\begin{pmatrix} \n \\ j \end{pmatrix}.
\end{gather}
Notice that $2\n-j\geq 1$ for $m \geq 4$ 
and the integral \refeq(sum-j) converges 
absolutely if $f$ is bounded with respect to $a$. 
\bglm If $m\geq 4$ is even, then we have 
\bqn \lbeq(co-kerb) 
G_0(\lam,x) =\sum_{j=0}^\n T_j^{(a)}
\left[ e^{i\lam|x|(1+2a)}\frac{(\lam|x|)^{j}}{|x|^{m-2}} \right]. 
\eqn 
\edlm 
\bgpf Write the Whittaker formula \refeq(co-ker) 
for $G_0(\lam, x)$ in the form    
\bqn \lbeq(co-ker-t) 
G_0(\lam,x) = \frac{e^{i\lam|x|}}
{C_m |x|^{m-2}}
\int^\infty_0 e^{-t}t^{\frac{m-3}{2}}
\left(t-2i\lam|x| \right)^{\frac{m-3}{2}}dt,
\eqn 
where $C_m= 2^{m-1}\pi^{\frac{m-1}{2}}
\Ga\left(\frac{m-1}{2}\right)$. In the integrand of 
\refeq(co-ker-t) we write 
\[
(t-2i\lam|x|)^{\frac{m-3}{2}}
=(t-2i\lam|x|)^{\n}(t-2i\lam|x|)^{-\frac{1}{2}}, 
\] 
expand  $(t-2i\lam|x|)^{\n}$ via the 
binomial formula and use the identity  
\bqn \lbeq(frac)
z^{-\frac12}= \frac1{\sqrt{\pi}}\int_0^\infty 
e^{-az}\, a^{-\frac12}\, da, \quad {\Re\, } z>0 
\eqn
for $(t-2i\lam|x|)^{-\frac12}$. Thus, the right hand side 
of \refeq(co-ker-t) becomes 
\[
\sum_{j=0}^\n \frac{(-2i)^j}{\sqrt{\pi}C_m}
\begin{pmatrix} \n \\ j \end{pmatrix} 
\iint_{\R_{+}^2} e^{-(1+a)t} t^{2\n-j}  
\left(e^{i\lam|x|(1+2a)}\frac{(\lam|x|)^{j}}{|x|^{m-2}} \right) 
\frac{dt}{\sqrt{t}}\frac{da}{\sqrt{a}}. 
\]
The integral converges absolutely if $m\geq 4$ and 
this is nothing but \refeq(co-kerb). \edpf 

\bglm For $m \geq 4$, $G_0(\lam,x)$ may also be written 
in the form 
\bqn 
G_0(\lam,x)= 
\sum_{j=0}^\n \frac{(\lam|x|)^{j}e^{i|x|\lam}}
{|x|^{m-2}}G_{j,m}(\lam|x|), 
\lbeq(coker-bc) 
\eqn 
where $G_{j,m}\in C^\infty(0,\infty)\cap C^{m-j-3}([0,\infty))$ 
and it satisfies for $\rho>1$ that 
\bqn  
|G_{j,m}^{(l)}(\rho)|\leq C_l \rho^{-l-\frac12}, \quad 
l=0,1, \dots.
\eqn 
\edlm  
\bgpf By changing the contour of integration, we have 
\begin{align*}
G_{j,m}(\rho)& = C_{jm}
\int_{0}^\infty e^{-2a\rho} 
(1+ia)^{-(2\n-j+\frac12)}
\frac{da}{\sqrt{a}} \\
& = \frac{C_{jm}}{\sqrt{\rho}}
\int_0^{\infty} e^{-2a} 
\left(1+\frac{ia}{\rho}\right)^{-(2\n-j+\frac12)} 
\frac{da}{\sqrt{a}}.
\end{align*}
The lemma is obvious from these expressions. 
\edpf 

\subsection{Spectral measure of $-\lap$}. 

Let $E_0(d\mu)$ be the spectral measure for 
$-\lap$. Then, with $\mu=\lam^2$,   
\[
E_0(d\mu)=\frac{1}{2\pi i}
(R_0(\mu+i0)-R_0(\mu-i0))d\mu  
=  \frac{1}{i\pi} (G_0(\lam)-G_0(-\lam)) \lam d\lam .
\]
We set $F(\lam)=\Phi(\lam^2)$ for $\lam\in \R$. 
Then, $F \in C_0^\infty(\R)$ and $F(|D|)= \Phi(H_0)$. 

\bglm \lblm(mulfo)  Let $m \geq 3$. 
Then, for $u, v \in \Sg(\R^m)$, both sides of the 
following equation 
can be continuously extended to $\lam=0$ and  
\bqn \lbeq(devi-by)
\lam^{-1} \la v, (G_0(\lam)-G_0(-\lam)) u \ra = 
\la |D|^{-1}v, (G_0(\lam)-G_0(-\lam)) u \ra , \quad \lam \geq 0.  
\eqn 
For continuous function $f$ of $\lam \in \R$ with 
compact support, we have for $\lam \geq 0$, 
\bqn 
\la v, G_0(\lam)u -G_0(-\lam))f(\lam)u \ra 
= \la v, (G_0(\lam)u -G_0(-\lam))f(|D|)u \ra \lbeq(fd-1) 
\eqn
\edlm 
\bgpf Let $\lam>0$. We have for any $u, v \in \Sg(\R^m)$ that  
\bqn 
\la v, (G_0(\lam)-G_0(-\lam)) u \ra= 
 \frac{\lam^{m-2}i}{2(2\pi)^{m-1} } \int_{\Si} 
\overline{\hat v(\lam\w)}\hat{u}(\lam\w)d\w  
\lbeq(spec-mea)
\eqn 
It follows, since 
$\widehat{|D|^{-1}v}(\lam\w)= \lam^{-1}\hat{v}(\lam\w)$, that 
\bqn 
\la |D|^{-1} v, (G_0(\lam)-G_0(-\lam)) u \ra
= \frac{\lam^{m-3}i}{2(2\pi)^{m-1} } 
\int_{\Si} \overline{\hat v(\lam\w)} \hat{u}(\lam\w)d\w.  
\lbeq(spec-mea-1)
\eqn 
The right hand side obviously extends to $u, v \in L^1(\R^m)$ 
to produce continuous functions of $\lam \geq 0$ when $m\geq 3$ 
and \refeq(devi-by) follows by comparing \refeq(spec-mea) 
and \refeq(spec-mea-1). If $f$ is continuous and bounded, 
then $\Fg f(|D|)u = f(|\xi|) \hat u(\xi)$ and 
\refeq(fd-1) likewise follows. 
\edpf 

We define the spherical average of a function $f$ on 
$\R^m$ by 
\begin{equation}\lbeq(M)
M(r, f) = \frac1{|\Si|}\int_{\Si} f(r\w) d\w, 
\quad \mbox{for all}\ r \in \R. 
\end{equation}
Here $\Si={\mathbb S}^{m-1}$ is the unit sphere 
and $|\Si|$ is its area. H\"older's inequality implies 
\bqn 
\left(\int_0^\infty |M(r)|^p r^{m-1}dr\right)^{1/p} 
\leq  \|f\|_p.
\eqn 
$M(r,f)$ is an even function of 
$r\in \R$. We then often use the following formula.  
For an even function $M(r)$ of $r\in \R$, 
define $\tilde M(\rho )$ by  
\bqn \lbeq(tiM)
\tilde M(\rho )= 
\int_{\rho }^\infty rM(r) dr 
\left(= -\int_{-\infty}^\rho  rM(r)dr\right).
\eqn 
\bglm Suppose $M(r)=M(-r)$ and $\la r \ra^2 M(r)$ is 
integrable. Then, 
\bqn  
\lbeq(MtildeM-1)
\int_{\R} e^{-ir\lam } r M(r)dr 
= \frac{\lam}{i} \int_\R e^{-ir \lam} \tilde M(r) dr,\ 
\int_{\R} \tilde{M}(r) dr= 
\int_{\R} r^2M (r) dr. 
\eqn 
\edlm 
\bgpf Since $rM(r)= -\tilde{M}(r)'$, integration by parts 
gives the first equation. We differentiate 
both sides of the first and set $\lam=0$. The second 
follows. \edpf 

We denote $\check u(x) =u(-x)$, $x\in \R^m$. 
(The sign $\check{\empty}$ will be reserved for this 
purpose and will not be used to denote the conjugate 
Fourier transform.)

\paragraph{Representation formula for odd dimensions.}
\begin{lemma}\lblm(p-stg) Let $m\geq 3$ be odd. 
Let $\p \in L^1(\R^m)$ and $u \in \Sg(\R^m)$. 
Let $c_j = |\Si|C_j$ where $C_j$ are the constants 
in {\rm \refeq(coker)}. Then, for $\lam>0$ 
we have 
\begin{multline} \lbeq(spheric)
\la \p\, |\, (G_0(\lam)-G_0(-\lam))u \ra \\ 
= \sum_{j=0}^{(m-3)/2} c_j (-1)^{j+1}\lam^{j} 
\int_{\R} e^{-i\lam r}r^{1+j} 
M(r,\overline{\p}\ast \check u)dr. 
\end{multline}
\end{lemma} 

\bgrm Combining \refeq(spec-mea) with \refeq(spheric) 
we have the identity: 
\[ 
\frac{\pi{i}\lam^{m-2}}{(2\pi)^m}\int_{\Si}
\overline{\hat{\p}(\lam\w)}\hat{u}(\lam\w)d\w
= \sum_{j=0}^{(m-3)/2} c_j(-1)^{j+1} \lam^{j} 
\int_{\R} e^{-i\lam r}r^{1+j} 
M(r,\overline{\p} \ast \check u)dr. 
\] 
This is particularly simple for $m=3$:
\bqn \lbeq(relation-3) 
\int_{\Si}
\overline{\hat{\p}(\lam\w)}\hat{u}(\lam\w)d\w
= \frac{8\pi^2{i}}{\lam} \int_{\R} e^{-i\lam r}r 
M(r,\overline{\p} \ast \check u)dr
\eqn 
\edrm 

\begin{proof} We compute $\la \p \,|\, G_0 (\lam )u \ra$ by 
inserting the integral kernel \refeq(coker) of $G_0(\lam)$. 
After changing the order of integration, we rewrite the 
integral by using polar coordinates. Then, 
\begin{align*}
\la \p \,|\, G_0 (\lam )u \ra 
= & \sum_{j=0}^{(m-3)/2}C_j
\int_{\R^m}\overline{\p(x)}
\left(\int_{\R^m}
\frac{\lam^je^{i\lam |y|}u(x-y)}{|y|^{m-2-j}}dy
\right)dx \\
=& \sum_{j=0}^{(m-3)/2} C_j 
\int_{\R^m}
\frac{\lam^j e^{i\lam |y|}(\overline{\p} \ast \check u)(y)}
{|y|^{m-2-j}}dy \\
=& \sum_{j=0}^{(m-3)/2} c_j 
\int_0^{\infty} \lam^j 
e^{i\lam r}r^{1+j} M(r,\overline{\p} \ast \check u)dr, \ 
c_j =C_j |\Si|. 
\end{align*}
Since $M(r)$ is an even function, change of variable 
$r$ to $-r$ yields  
\[
-\la \p\,|\, G_0 (-\lam)u \ra =
\sum_{j=0}^{(m-3)/2} c_j 
\int_{-\infty}^0 \lam^j e^{i\lam r}r^{1+j} 
M(r,\overline{\p} \ast \check u)dr. 
\]
Add both sides of last two equations and change 
the variable $r$ to $-r$. This produces \refeq(spheric). 
\end{proof} 

\paragraph{Spectral measure in even dimensions.} 
If $m$ is even, we have the analogue of 
\refeq(spheric). For a function $A(r)$ on $\R$ and $a>0$, 
define   
\[
A^a(r)=A(r/(1+2a))
\] 
and write  
$M_{\overline{\p } \ast \check{u}}(r)$ 
for $M(r, \overline{\p } \ast \check{u})$ 
for shorting the formula
. 

\begin{lemma} Let $m\geq 2$. Let 
$\p  \in L^1 (\R^m)$ and 
$u \in {\Sg}(\R^m)$. Then
\begin{align} 
& \langle \p ,(G_0(\lam)- G_0(-\lam))u\rangle 
= \sum_{j=0}^\n (-1)^{j+1}|\Si| T_j^{(a)} \left[
\frac{\lam^j 
\Fg (r^{j+1} M^{a}_{\overline{\p } \ast \check{u}})(\lam)}
{(1+2a)^{j+2}}
\right]  \lbeq(usformula) \\
& \qquad 
= \sum_{j=0}^\n (-1)^{j+1}|\Si| T_j^{(a)} \left[
\lam^j 
\Fg (r^{j+1} M_{\overline{\p } \ast \check{u}})((1+2a)\lam)
\right]. \lbeq(usformula-11)
\end{align}
The term with $j=0$ in the right of 
\refeq(usformula) may also be written as 
\bqn \lbeq(usformula-add)
i|\Si| T_0^{(a)}\left[\frac{\lam  
(\Fg \widetilde{M^a_{\overline{\p } \ast \check{u}}})(\lam)}
{(1+2a)^2} \right].
\eqn  
\end{lemma}  
\bgpf  Define 
$ A_j(\lam,r,a)= e^{i\lam r(1+2a)}(\lam r)^{j}r^{-(m-2)}$  
and operator $A_j(\lam,a)$ by 
\[ 
A_j(\lam,a)u(x)= \int_{\R^m}
A_j(\lam, |y|, a)u(x-y) dy, \quad j=0, \dots, \n.
\]
Then, \refeq(co-kerb) and change of the order of integration 
imply  
\bqn \lbeq(gom) 
\la \p , (G_0(\lam)-G_0(-\lam)) u \ra  =\sum_{j=0}^\n T_j^{(a)}
\left[ \la \p , (A_j(\lam,a)-A_j(-\lam,a))u \ra \right].
\eqn 
We have, as in odd dimensions, that 
for $u \in \Sg(\R^m)$ and $\p  \in L^1(\R^m)$ 
\begin{align*}
& \la \p , A_j(\lam,a) u\ra= \int_{\R^{m}}\left(\int_{\R^m}
\overline{\p (x)} A_j(\lam, |y|, a)u(x-y)dy \right)dx \notag \\
& = \int_{\R^m} A_j(\lam, |y|,a) 
(\overline{\p } \ast \check{u})(y)dy 
= |\Si|\int_0^\infty e^{i(1+2a)\lam{r}}(\lam{r})^j 
r M_{\overline{\p } \ast \check{u}}(r) dr. 
\end{align*} 
Replacing $\lam$ to $-\lam$ and changing the variable 
$r$ to $-r$, we have 
\[
-\la \p , A_j(-\lam,a) u\ra
= |\Si| \int_{-\infty}^0 e^{i(1+2a)\lam{r}}(\lam{r})^j r 
M_{\overline{\p } \ast \check{u}}(r) dr,
\] 
where we used that 
$M_{\overline{\p } \ast \check{u}}(-r)
=M_{\overline{\p } \ast \check{u}}(r)$.
Adding these two yields
\bqn \lbeq(k-3)
\la \p , (A_j(\lam,a)- A_j(-\lam,a)) u\ra 
= |\Si| \int_{\R} e^{i(1+2a)\lam r}(\lam r)^j r 
M_{\overline{\p } \ast \check{u}}(r) dr .
\eqn 
Changing $r$ to $-r$ in the right of \refeq(k-3) and 
plugging the result with \refeq(gom), 
we obtain \refeq(usformula-11). If we change the variable 
$r$ to $-r/(1+2a)$, \refeq(k-3) becomes   
\[ 
\frac{(-1)^{j+1}|\Si|}{(1+2a)^{j+2}}
\int_{\R} e^{-i\lam r} \lam^j r^{j+1} 
M^{a}_{\overline{\p } \ast \check{u}}(r) dr 
= \frac{(-1)^{j+1}|\Si| \lam^j}{(1+2a)^{j+2}}
\Fg (r^{j+1} M^{a}_{\overline{\p } \ast \check{u}})(\lam).
\]
If we substitute this for 
$\la \p , (A_j(\lam,a)-A_j(-\lam,a))u\ra$ 
in \refeq(gom), \refeq(usformula) follows. 
If we use the second equation of \refeq(MtildeM-1)  
the right of the last equation for $j=0$ becomes   
\[
\frac{i\lam(\Fg \widetilde{M_{\overline{\p } 
\ast \check{u}}^a})(\lam)}
{(1+2a)^2}|\Si|
\]
and \refeq(usformula-add) follows. 
\end{proof}

\subsection{Proof of \refthb(theo) for odd $m\geq 3$.} 

We substitute \refeq(res), \refeq(res-5) and \refeq(res-7) 
for $S(\lam)$ in the equation \refeq(zs) respectively 
for $m=3,5$ and $m\geq 7$. We write $\{\f_1, \dots, \f_d\}$ 
for the orthonormal basis of the $0$ eigenspace $\Eg$ of $H$ 
and define 
\bqn \lbeq(ws2-7)
Z_{s1}u  = \sum_{j=1}^d \frac{i}{\pi} \int_0^\infty 
G_0(\lam)|V\f_l\ra \la V\f_l|(G_0(\lam)-G_0(-\lam))
F(\lam)\lam^{-1}u d\lam. 
\eqn 
Then, we have $Z_s = Z_{s1}$ if $m \geq 7$ and, if 
$m=3$ and $m=5$, we have less singular extra term 
$Z_{s0}$ given below and $Z_s= Z_{s0}+ Z_{s1}$:  
For $m=5$, with $\ph= P_0 V$ 
\[
Z_{s0}u= \int_0^\infty G_0(\lam)|\, V\ph \ra 
\la V\ph \, |\,(G_0(\lam)-G_0(-\lam))u\ra F(\lam)d\lam
\] 
and for $m=3$, with $\f_0$ being the canonical resonance 
\[
Z_{s0}u= \sum_{l,n=0}^d  a_{ln}
\int_0^\infty G_0(\lam)|\, V\f_l \ra 
\la V\f_n\, |\,(G_0(\lam)-G_0(-\lam))u\ra F(\lam)d\lam
\] 
where $a_{ln}, \ 0\leq l,n\leq d$ are constants 
such that $a_{0l}=a_{l0}=0$ for $1\leq l \leq d$. 
Thus, for proving \refth(theo) for odd $m\geq 3$, 
it suffices to prove the following proposition. 

\bgprp \lbprp(prp) 
Let $1<p<3$ if $m=3$ and  $1<p<m/2$ if $m\geq 5$.
\ben
\item[{\rm (1)}] For any $\f, \p \in \Ng$, 
the operator $\tilde{Z}_{s0}(\f,\p)$ defined by 
\bqn 
\tilde{Z}_{s0}(\f,\p)u = \frac{i}{\pi} \int_0^\infty 
G_0(\lam)V\f \ra \la V\p |(G_0(\lam)-G_0(-\lam))
F(\lam)u d\lam \lbeq(ws-1) 
\eqn 
is bounded in $L^p(\R^m)$ if $m=3, 5$. 
\item[{\rm (2)}] For any $\f, \p \in \Eg$, the operator
$\tilde{Z}_{s1}(\f,\p)$ defined by 
\bqn 
\tilde{Z}_{s1}(\f,\p)u = \frac{i}{\pi} \int_0^\infty 
G_0(\lam)V\f \ra \la V\p |(G_0(\lam)-G_0(-\lam))
F(\lam)u \frac{d\lam}{\lam}. \lbeq(ws-2)
\eqn 
is bounded in $L^p(\R^m)$ for all odd $m\geq 3$. 
\een
\edprp
\bgpf Write $\tilde{Z}_{s\ell}$ for 
$\tilde{Z}_{s\ell}(\f,\p)$, $\ell=0,1$ for simplicity. 
We replace $G_0(\lam)$ by \refeq(coker) and use 
\refeq(spheric) for $\la V\p,(G_0(\lam)-G_0(-\lam))u\ra$. 
Writing $M(r)$ for $M(r, V\p \ast u)$, we obtain  
\bqn 
\tilde{Z}_{s\ell}u= \frac{i}{\pi}
\sum_{j,k=0}^{\frac{m-3}2} 
(-1)^{j+1}C_k c_j Z_{s\ell}^{(j,k)}u, 
\lbeq(expre-0)
\eqn 
where $Z_{s\ell}^{(j,k)}$, $0 \leq j,k \leq \frac{m-3}2$, 
are given by  
\begin{gather} 
Z_{s\ell}^{(j,k)}u(x)= \int_{\R^m} 
\frac{V\f (y)}{|x-y|^{m-2-k}}
K^{(j,k)}_\ell(|x-y|)dy, 
\lbeq(ws1jk)
\\ 
K^{(j,k)}_\ell(\rho )= \frac1{2\pi}
\int_0^\infty e^{i\lam\rho }\lam^{j+k-\ell}
\left(
\int_{\R} e^{-i\lam r} r^{j+1} M(r)dr
\right)F(\lam)d\lam. \lbeq(Kjk-def)
\end{gather}
By using Young's inequality and polar coordinates, 
we have from \refeq(ws1jk) that 
\bqn \lbeq(Young)
\|Z_{s\ell}^{(j,k)} u\|_p \leq C \|V\f\|_1 
\left(\int_0^\infty \frac{|K_\ell^{(j,k)}(\rho)|^p}
{\rho^{(m-2-k)p}}
\rho^{m-1}d\rho 
\right)^{1/p}. 
\eqn
This will often be the starting point of 
estimates though some modification will be in some cases. 
When $\ell=0$, we have by virtue of \reflms(add-cut,max) 
that 
\bqn \lbeq(k0rx)
K_{0}^{(j,k)}(\rho ) 
= \{(\Fg^\ast (\lam^{j+k}F)\ast\Hg(r^{j+1}M(r))\}(\rho)
\absleq C \Mg(r^{j+1}M)(\rho)
\eqn
for all $0\leq j,k \leq \frac{m-3}2$. 
The proposition is proved by the series of lemmas 
given below. \edpf 

\bglm \lblm(1) 
Let $m=3$ and $1<p<3$. Then,  
$\|Z_{s0} u\|_p \leq C\|u\|_p$ for a constant $C$ 
independent of $u\in C_0^\infty(\R^3)$.
\edlm 
\bgpf If $m=3$, we have $j=k=0$ only and we omit the 
suffix $(j,k)$ from $Z_{s0}^{(j,k)}u$ and $K_0^{(j,k)}$. 
Let $3/2<p<3$ first. Then,  
$\rho^{2-p}$ is an $A_p$ weight on $\R$ and, applying 
\reflm(ap) to \refeq(k0rx), we obtain  
\begin{align}
\left(\int_0^\infty 
|K_{0}(\rho)|^p \rho^{2-p} d\rho\right)^{1/p}  
& \leq C \left(\int_0^\infty M(r)^p r^2 dr\right)^{1/p} \\
& \leq C \|V\p \ast u\|_p \leq C \|V\p\|_1 \|u\|_p. 
\lbeq(3d-b1)
\end{align}
The lemma for $3/2<p<3$ follows from this and \refeq(Young).
For $1<p<3/2$, we apply integration by parts and apply 
\reflms(add-cut,max) to the resulting equation and obtain   
\begin{align}
K_0(\rho )& =
\frac{i}{2\pi\rho }\int_0^\infty e^{i\lam\rho } 
\left(F(\lam)\int_\R e^{-ir\lam} rM (r) dr \right)' d\lam 
\notag \\
& \absleq C \rho^{-1}(\Mg(r^2 M)(\rho ) 
+\Mg(r M)(\rho )).
\lbeq(3d-3)
\end{align}
If $1<p<3/2$, then, $\rho^{2-2p}$ is an $A_p$ weight on 
$\R$ and \refeq(3d-3) and \reflm(ap) imply  
\begin{align}
& \left(\int_0^\infty 
|K_{0}(\rho)|^p \rho^{2-p} d\rho\right)^{1/p} 
\leq \left(\int_0^\infty 
|\Mg(r^2 M)(\rho )+ \Mg(r M)(\rho )|^p
\rho^{2-2p}d\rho\right)^{1/p} \notag \\
& \leq C \left(\int_0^\infty |M(r)|^p r^2 dr\right)^{1/p} 
+ C \left(\int_0^\infty 
|M(r)|^p r^{2-p}dr\right)^{1/p}. 
\lbeq(3d-c)
\end{align} 
We estimate the first on the right of \refeq(3d-c) 
as in \refeq(3d-b1) and the second by 
\begin{align}
& \left(\int_0^\infty 
|M(r)|^p r^{2}dr\right)^{1/p}
+ \left(\int_0^1
|M(r)|^p r^{2-p}dr\right)^{1/p} \notag \\
& \leq C (\|V\p \ast u\|_p + \|V\p\ast u\|_\infty) 
\leq C (\|V\p\|_1+ \|V\p\|_{p'})\|u\|_p, \lbeq(3d-g)
\end{align}
where $p'=p/(p-1)$ is the dual exponent of $p$ 
and we used obvious estimate 
\bqn  
\sup |M(r)| \leq \|V\p \ast u\|_\infty.  \lbeq(supM)
\eqn 
Combining \refeq(3d-g) with \refeq(Young) we have  
the lemma for $1<p<3/2$ as well. Then, the interpolation 
theorem (\cite{BL}) completes the proof of the lemma. 
\edpf 

\bglm \lblm(2)
Let $m=5$ and $1<p<5/2$. Then,  
$\|\tilde{Z}_{s0} u\|_p \leq C\|u\|_p$ for a constant $C$ 
independent of $u\in C_0^\infty(\R^5)$.
\edlm 
\bgpf By virtue of the interpolation theorem, it suffices 
to prove the lemma for $1<p<5/4$ and for $2\leq p<5/2$.  
Though $m=5$ in this proof, we will also write $m$ 
for the space dimension $5$ as this is sometimes 
conveninet. We integrate by parts $k+1$-times 
in the right of \refeq(Kjk-def) for $\ell=0$. Since all 
derivatives up to the order $k$ of 
$\lam^{j+k}\left(
\int_{\R} e^{-i\lam r} r^{j+1} M(r)dr \right)F(\lam)$ 
vanish at $\lam=0$, no boundary terms appear and 
after estimating the resulting equation by using 
\reflms(add-cut,max), we obtain in addition to \refeq(k0rx) 
that 
\begin{align}
K^{(j,k)}_0(\rho )& = \frac{i^{k+1}}{2\pi\rho^{k+1}}
\int_0^\infty e^{i\lam\rho }
\left(\frac{\pa}{\pa\lam}\right)^{k+1}
\left(\lam^{j+k}F(\lam) 
\int_{\R} e^{-i\lam r} r^{j+1} M(r)dr
\right)d\lam \notag \\
& \absleq \frac{C}{\rho^{k+1}}\sum_{l=0}^{k+1} 
\Mg \Hg (r^{j+l+1} M(r))(\rho).  \lbeq(5-a)
\end{align} 
Let $1<p<m/(m-1)$. Then, $|r|^{-(p-1)(m-1)}$ is an 
$A_p$ weight on $\R$ and 
\refeq(5-a) and \reflm(ap) imply that the integral of 
on the right of \refeq(Young) is bounded by a constant times 
\begin{align}
& \sum_{l=0}^{k+1}
\left(\int_0^\infty \frac{|\Mg\Hg(r^{j+l+1}M)(\rho)|^p}
{\rho^{(p-1)(m-1)}} d\rho\right)^{1/p} 
\leq C \sum_{l=0}^{k+1}
\left(\int_0^\infty \frac{|r^{j+l+1}M(r)|^p}
{r^{(p-1)(m-1)}}dr \right)^{1/p} \notag \\
& \leq C \left(
\int_{\R} 
\frac{|(|r|+|r|^{m-1})M(r)|^p}{r^{(p-1)(m-1)}}dr 
\right)^{1/p} 
\leq C (\|V\p\|_1+ \|V\p\|_{p'})\|u\|_p . \lbeq(5d-d)
\end{align}
Here the last estimate is obtained as in \refeq(3d-g) 
by using \refeq(supM). This proves the lemma 
for $1<p<5/4$. 

If $2 \leq p<m/2$, then $|r|^{m-1-2p}$ is an $A_p$ weight on 
$\R$ and $m-2-k=2$ if $m=5$ and $k=1$. Hence, 
estimating the right of \refeq(Young) by using \refeq(k0rx) 
and \reflm(ap) as previously, we obtain   
\begin{align}
\|Z^{(j,1)}_0 u\|_p & \leq 
C \|V\f\|_1 \left(
\int_0^\infty 
|\Mg \Hg (r^{j+1} M(r))|^p \rho^{m-1-2p} d\rho\right)^{1/p} 
\notag 
\\
& \leq C 
\|V\f\|_1 \left(
\int_0^\infty |M(r))|^p \rho^{m-1-(1-j)p} d\rho\right)^{1/p} 
\leq C \|u\|_p.  \lbeq(5d-e)
\end{align}
When $k=0$, we split the integral in \refeq(ws1jk) and 
estimate   
\begin{multline*} 
Z_{s0}^{(j,0)}u(x) 
\absleq 
\int_{|x-y|\leq 1} 
\frac{|V\f(y)|}{|x-y|^{3}}
|K^{(j,0)}_0 (|x-y|)|dy \\
+\int_{|x-y|\geq 1} 
\frac{|V\f(y)|}{|x-y|^{2}}
|K^{(j,0)}_0 (|x-y|)|dy = I_1(x) + I_2(x). 
\end{multline*}
We have by using Young's inequality that 
\bqn 
\|I_2\|_p \leq \|V\f\|_1 
\left\|\frac{K^{(j,0)}_0(|x|)}{|x|^2}\right\|_p 
\lbeq(efg-0)
\eqn 
and, applying \reflm(ap) to \refeq(k0rx) again,     
\begin{align} 
& \left\|\frac{K^{(j,0)}_0(|x|)}{|x|^2}\right\|_p 
= C \left(\int_0^\infty 
|K^{(j,0)}_0(\rho)|^p \rho^{m-1-2p}d\rho\right)^{1/p} 
\notag \\
& \leq C \left(\int |M(r)|^p \rho^{m-1-(1-j)p}d\rho\right)^{1/p}
\leq C \|u\|_p. \lbeq(efg) 
\end{align} 
For $I_1(x)$ we first use H\"older's inequality and 
then apply \refeq(efg) to the second factor of the result 
to obtain 
\begin{align*}
|I_1(x)| & \leq \left(\int_{|y|\leq 1} 
\frac{|V\f(x-y)|^{p'}}{|y|^{p'}}dy \right)^{1/{p'}} 
\left\|\frac{K^{(j,0)}_0(|x|)}{|x|^2}\right\|_p \\
& \leq \left(\int_{|y|\leq 1} 
\frac{|V\f(x-y)|^{p'}}{|y|^{p'}}dy \right)^{1/{p'}} \|u\|_p,
\end{align*} 
where ${p'}$ is the dual exponent of $p$ and 
$\frac{m}{m-2}<{p'}\leq 2 \leq p<m/2$.  Hence Minkowski's  
inequality implies 
\bqn \lbeq(efg-2)
\|I_1 \|_p \leq C \|u\|_p \||V\f|^{p'}\|_{p/{p'}}^{1/{p'}} 
\left(\int_{|y|\leq 1} |y|^{-{p'}} dy \right)^{1/{p'}}
= C \|V\f\|_p \|u\|_p .
\eqn 
Combining \refeq(efg-0), \refeq(efg) and \refeq(efg-2),  
we obtain the lemma also for $2\leq p<m/2$.  \edpf 

We next study $\tilde{Z}_{s1}u$ and prove the 
second statement of the proposition. Thus, $\f, \p \in \Eg$ 
in what follows. We begin with the case $m=3$. 
\bglm Let $m=3$ and $1<p<3$. Then, for a constant $C_p$ 
independent of $u\in C_0^\infty(\R^3)$ we have 
$\|\tilde{Z}_{s1}u\|_p \leq C_p \|u\|_p$. 
\edlm 
\bgpf Define $\eta(x) =|D|^{-1} (V\p)$. 
Then, by virtue of \refeq(devi-by), we have that   
\[
\tilde{Z}_{s1}u = \int_0^\infty 
G_0(\lam)|V\f\ra 
\la \eta\, |\, (G_0(\lam) -G_0(-\lam))u\ra F(\lam)d\lam 
\]
which is the same as $\tilde{Z}_{s0} u$ with 
$V\p$ being replaced by $\eta$. Thus, 
if we define $M(r)=M(r, \eta \ast \check{u})$, 
then the proof of \refeq(3d-b1), \refeq(3d-c) and 
\refeq(3d-g) implies that 
\bqn \lbeq(3d-h)
\|\tilde{Z}_{s1}u\|_p \leq C 
\left(\int_{0}^\infty |M(r)|^p r^2 dr\right)^{1/p} 
\leq C \|\eta \ast u\|_p 
\eqn  
when $3/2<p<3$ and, when $1<p<3/2$ that  
\bqn \lbeq(3d-i)
\|\tilde{Z}_{s1}u\|_p \leq C 
\left(\int_{0}^\infty |M(r)|^p r^{2-p} dr\right)^{1/p} 
\leq C (\|\eta \ast u\|_p + \|\eta \ast u\|_\infty)
\eqn  
The operator $|D|^{-1}$ is the convolution with 
$C|x|^{-2}$ with a constant $C$ and $\p$ satisfies  
$|V(x) \p(x) |\leq C \ax^{-\delta-2}$ and 
$\int V\p dx=0$. It follows that $\eta(x)$ is  
bounded and as $|x|\to \infty$, 
\begin{align*}
\eta(x) & = C\int 
\left(\frac1{|x-y|^2}-\frac1{|x^2|}\right)
(V\p)(y)dy \\
& = C\int 
\frac{2x\cdot y -|y|^2}{|x|^2|x-y|^2} (V\p)(y) dy  
\ds = \sum_{k=1}^3 \frac{C_{jk}x_k}{|x|^4} 
+ O(|x|^{-4}).  
\end{align*}
Thus $\eta \in L^q(\R^3)$ for any $1<q \leq \infty$ and 
the convolution with $\eta(x)$ is  
bounded in $L^p$ for any $1<p<\infty$ by the 
Calder\`on-Zygmund theory 
(see e.g. \cite{Stein}, pp. 30-36).
Thus, the right hand sides of \refeq(3d-h) and 
\refeq(3d-i) are both bounded by $C \|u\|_p$ 
for $3/2<p<3$ and $1<p<3/2$ respectively. 
\edpf 

\bglm Let $m \geq 5$ be odd and $1<p<\frac{m}{m-1}$. 
Then, for a constant $C_p$ independent of 
$u\in C_0^\infty(\R^m)$, we have 
$\|\tilde{Z}_{s1}u\|_p \leq C \|u\|_p$. 
\edlm 
\bgpf Here again $M(r)=M(r, (V\p)\ast \check u)$. 
We first prove 
\bqn \lbeq(asp-1)
\|\tilde{Z}_{s1}^{(jk)}u\|_p \leq C \|u\|_p, 
\quad 2\leq j \leq \tfrac{m-3}2, \ 
0\leq k \leq \tfrac{m-3}2. 
\eqn 
The proof is almost a repetition of that for $m=5$. 
If $2\leq j \leq \tfrac{m-3}2$, then 
all derivatives of order up to $k$ of 
$\lam^{j+k-1}F(\lam)\int_{\R} e^{-i\lam r} r^{j+1} M(r)dr$ 
vanish at $\lam=0$  and $k+1$ integrations by parts 
show as in \refeq(5-a) that  
\bqn 
K^{(j,k)}_1 (\rho) \absleq 
\sum_{l=0}^{k+1}\frac{C_{jkl}}{\rho^{k+1}}
\Mg\Hg(r^{j+1+l}M)(\rho).
\eqn 
Since $r^{-(m-1)(p-1)}$ is an $A_p$ weight on 
$\R$ for $1<p<\frac{m}{m-1}$, we have   
\begin{align}  
& \left(\int_0^\infty \frac{|K^{(j,k)}_1(\rho)|^p}
{\rho^{(m-2-k)p}}\rho^{m-1} d\rho\right)^{1/p} 
\leq C \sum_{l=0}^{k+1} 
\left(\int_0^\infty \frac{|M(r)|^p}
{r^{(m-2-j-l)p}}r^{m-1} dr \right)^{1/p} \notag 
\\
& \leq 
C \left(\int_0^\infty |M(r)|^p r^{m-1}dr \right)^{1/p}
+ C \left(\int_0^1 \frac{|M(r)|^p}{r^{(m-4)p}} 
r^{m-1}dr \right)^{1/p} \notag \\
& \leq C (\|(V\p)\ast u\|_p  + 
\|(V\p)\ast u\|_\infty)  
\leq C(\|V\p\|_1 + \|V\p\|_{p'})\|u\|^p_p.
\lbeq(aspreviously)
\end{align}
Here at the last step we used \refeq(supM) and that 
$0\leq p(m-4)<m$. We next prove that  
\bqn \lbeq(joint) 
\|(c_0 Z_{s1}^{(0,k)}-c_1 Z_{s1}^{(1,k)})u\|_p \leq C \|u\|_p, 
\quad 0 \leq k \leq \tfrac{m-3}2.
\eqn 
This and \refeq(aspreviously) will prove the lemma. 
Recall $c_j= C_j|\Si|$, $C_j$ are constants 
of \refeq(coker) and $C_0-i C_1=0$. 
Using \refeq(MtildeM-1), we rewrite   
\bqn 
K^{(0,k)}_1(\rho)=\frac{-i}{2\pi} \int^\infty_0 
e^{i\lam\rho}\lam^{k}F(\lam)
\left(\int_{\R} e^{-i\lam r}\tilde M(r)dr\right)d\lam .
\lbeq(K00)
\eqn
Since the derivatives 
up to the order $k-1$ of $\lam^{k}F(\lam)
\left(\int_{\R} e^{-i\lam r}\tilde M(r)dr\right)$
vanishes at $\lam=0$ and 
\[
\frac{\pa^k}{\pa \lam^k}
\left. \left(
\lam^{k}F(\lam)
\int_{\R} e^{-i\lam r}\tilde M(r)dr\right)
\right\vert_{\lam=0} = k! \int_{\R}r^2 M(r)dr,  
\]
integration by parts using 
$(\pa/\pa\lam)^{k+1}e^{i\lam\rho}=(i\rho)^{k+1}e^{i\lam\rho}$ 
yields that   
\begin{align}
& K^{(0,k)}_1(\rho) =\frac{i^k}{2\pi\rho^{k+1}}
\left(k! \int_{\R}r^2 M(r)dr \right. \notag \\
& \quad \left. + 
\sum_{l=0}^{k+1}
\begin{pmatrix} k+1 \\ l \end{pmatrix}
\int_0^\infty e^{i\lam\rho}
(\lam^{k}F)^{(k+1-l)} 
\int_{\R} e^{-i\lam r}(-ir)^{l} \tilde{M}dr
d\lam \right) .  \lbeq(K0k) 
\end{align}
Applying likewise integration by parts to $K_{1k,1}(\rho)$, 
we obtain 
\begin{align}
& K^{(1,k)}_1(\rho)=\frac1{2\pi} \int^\infty_0 
e^{i\lam\rho}\lam^{k}F(\lam)
\left(\int_{\R} e^{-i\lam r}r^2 M(r)dr\right)d\lam \notag 
\\
&= \frac{i^k}{2\pi{i}\rho^{k+1}}
\left(-k! \int_{\R}r^2 M(r)dr \right. 
\notag 
\\
& \left. - 
\sum_{l=0}^{k+1}
\begin{pmatrix} k+1 \\ l \end{pmatrix}
\int_0^\infty e^{i\lam\rho}
(\lam^{k}F)^{(k+1-l)} 
\int_{\R} e^{-i\lam r}(-ir)^{l}r^2 M dr
d\lam \right) . \lbeq(K1k)
\end{align}
Since $c_0-ic_1=0$, we have by combining 
\refeq(K0k) and \refeq(K1k) that 
\begin{align}
& c_0 K^{(0,k)}_1(\rho) - c_1 K^{(1,k)}_1(\rho) 
= \frac{c_0 i^{k-l}}{\rho^{k+1}}
\sum_{l=0}^{k+1}
\begin{pmatrix} k+1 \\ l \end{pmatrix} \\
& \left(\{\Fg^\ast((\lam^{k}F)^{(k+1-l)})\ast  
(\Hg(r^{l}\tilde{M})\}(\rho) 
+ \{\Fg^\ast((\lam^{k}F)^{(k+1-l)})\ast  
\Hg(r^{l+2} M)\}(\rho) \right)  \notag \\ 
& \absleq 
\frac{C}{\rho^{k+1}}
\sum_{l=0}^{k+1}(\Mg \Hg(r^{l}\tilde{M})(\rho)+ 
\Mg \Hg(r^{l+2} M)(\rho)). \lbeq(coK)  
\end{align}
It follows again by using Young's inequality 
and the weighted inequality that    
\begin{align} 
& \|(c_0 Z_{s1}^{(0,k)}-c_1 Z^{(1,k)}_{s1})u\|_p \notag \\
& \leq  
C \sum_{l=0}^{k+1} \|V\f\|_1  
\left(\int_0^\infty 
\frac{|\Mg \Hg(r^{l}\tilde{M}(\rho)|^p + 
|\Mg \Hg(r^{l+2}M(\rho)|^p}
{\rho^{(m-1)p}} \rho^{m-1}d\rho\right)^{1/p} \notag \\
& \leq C \sum_{l=0}^{k+1} \|V\f\|_1 
\left(\int_0^\infty 
(|\tilde M(r)|^p r^{pl} + 
|M(r)|^p r^{p(l+2)}) r^{m-1-p(m-1)}dr\right)^{1/p} \notag \\
& \leq 
C\sum_{l=0}^{k+1} \|V\f\|_1
\left(\int_0^\infty |M(r)|^p r^{p(l-m+3)+m-1}dr\right)^{1/p},  
\lbeq(W0ka)
\end{align}
where we used Hardy's inequality at the last step,  
remembering the definition \refeq(tiM).   
Here $ -m+3\leq l-m+3 \leq 0$ when $0\leq l \leq k+1$ and  
\begin{align*}
& \refeq(W0ka) \leq C\|V\f\|_1 
\left(\int_0^1 
|M(r)|^p r^{m-1-p(m-3)}dr 
+ \int_0^\infty |M(r)|^p r^{m-1} dr 
\right)^{1/p} \\ 
& \leq C \|V\f\|_1 
(\|V\p \ast u\|_\infty + \|V\p \ast u\|_p)
\leq C \|V\f\|_1 (\|V\p\|_{p'}+ \|V\p\|_p)\|u\|_p. 
\end{align*}
This completes the proof of the lemma. 
\edpf

The following lemma completes the proof of \refprp(prp). 
\bglm \lblm(lastlm-odd)
Let $m \geq 5$ be odd and $\max(2,m/3)<p<m/2$. 
Then, for a constant $C_p$ independent of 
$u\in C_0^\infty(\R^m)$, we have 
 $\|\tilde{Z}_{s1} u\|_p \leq  C_p \|u\|_p$.
\edlm 
\bgpf  We prove for $0 \leq j,k \leq \frac{m-3}2$ that 
\bqn \lbeq(lm2-2)
\|Z_{s1}^{(j,k)}u \|_p \leq C \|u\|_p, \quad 
\max(2,m/3)<p<m/2.
\eqn  
Here again $M(r)=M(r, V\p \ast u)$. We have 
\bqn \lbeq(split-odd)
Z_{s1}^{(j,k)}u(x) 
= \left(\int_{|y|<1} + \int_{|y|\geq 1}\right) 
\frac{V\f(x-y)}{|y|^{m-2-k}}
K_{jk,1}(|y|)dy =I_1(x)+ I_2(x).
\eqn 
Let $j \geq 1$ first. We have 
\bqn \lbeq(odd-ip)
K^{(j,k)}_1(r)=\frac1{2\pi}
\int_0^\infty e^{i\lam\rho}\lam^{j+k-1}F(\lam) 
\left(\frac{i\pa}{\pa\lam}\right)^{j-1}
\left(\int_{\R} e^{-i\lam r} r^{2} M(r)dr\right)
d\lam .
\eqn 
We then apply integration by parts in a way  
sightly differently from that used in the proof of the 
previous lemma and obtain  
\begin{align}
K^{(j,k)}_1 (r)   
& =\sum_{l=0}^{j-1} \begin{pmatrix} j-1 \\ l \end{pmatrix}
\rho^l \{\Fg^\ast((\lam^{j+k-1}F(\lam))^{(j-1-l)})\ast 
\Hg(r^{2} M)\}(\rho) \notag  
\\
& \absleq \sum_{l=0}^{j-1} C_{jl}\rho^{l} 
\Mg \Hg (r^2 M)(\rho)  \quad j \geq 1. \lbeq(smallr)
\end{align}
When $j=0$, we use $\tilde M(r)$ of \refeq(tiM) 
and estimate as   
\bqn 
K^{(0,k)}_1 (\rho) =\frac1{2i\pi}
\int_0^\infty e^{i\lam\rho}\lam^{k}F(\lam)
\left(\int_{\R} e^{-i\lam r} \tilde{M}(r)dr\right)
d\lam \absleq \Mg \Hg(\tilde{M})(\rho). \lbeq(smallr-af)
\eqn 
Let $j\geq 1$. If $\rho\geq 1$, then \refeq(smallr) is bounded by 
$C \rho^{j-1} \Mg \Hg (r^2 M)(\rho)$ and  
$m-k-j-1\geq 2$. It follows by using 
Young's inequality and the weighted inequality that 
\begin{align}
\|I_2\|_p 
& \leq \|V\f\|_1 
\left(\int^\infty_1 
\frac{|K^{(j,k)}_1|^p}
{\rho^{p(m-2-k)}}\rho^{m-1}d\rho \right)^{1/p} \lbeq(cde-0) 
\\
& \leq C \|V\f\|_1 
\left(\int^\infty_1 
\frac{|\Mg \Hg (r^2 M)(\rho)|^p}
{\rho^{2p}}\rho^{m-1}d\rho \right)^{1/p} \notag \\ 
& \leq C \|V\f\|_1 \left(\int_\R |M(r)|^p r^{m-1}dr\right)^{1/p} 
\leq C \|V\f\|_1^p \|u\|_p^p, 
\lbeq(cde)
\end{align}
for $|r|^{m-1-2p}$ is an $A_p$ weight on $\R$ for $m/3<p<m/2$. 
When $j=0$, we estimate the right of \refeq(cde-0) 
by first using \refeq(smallr-af), remarking 
$m-2-k\geq 2$ for $0\leq k \leq\frac{m-3}2$, 
apply \reflm(ap) and then Hardy's inequality 
consecutively. Then,    
\begin{multline*}
\|I_2\|_p \leq \|V\f\|_1 
\left(\int^\infty_1 \frac{|\Mg \Hg (\tilde{M})(\rho)|^p}
{\rho^{2p}}\rho^{m-1}d\rho\right)^{1/p} \\
\leq C \|V\f\|_1 
\left(\int_\R |\tilde{M}|^p r^{m-1-2p}dr\right)^{1/p} 
\leq 
C \|V\f\|_1\left(\int_\R |{M}(r)|^p r^{m-1}dr\right)^{1/p}. 
\end{multline*}
The right side is bounded by $C\|V\f\|_1 \|V\p\|_1 \|u\|_p$ 
as previously. This and \refeq(cde) show 
$\|I_2\|_p \leq C \|u\|_p$. 
We next estimate $\|I_1\|_p$. By virtue of 
H\"older's inequality 
\bqn 
|I_1(x)| \leq \left(\int_{|y|<1}\left|\frac{V\f(x-y)}
{|y|^{m-4-k}}\right|^{p'}dy \right)^{1/p'}
\left(\int_{|y|<1}\left|\frac{K_{jk}(|y|)}
{|y|^{2}}\right|^p dy \right)^{1/p}. \lbeq(smallr-b)
\eqn 
When $0<\rho\leq 1$, 
\refeq(smallr) and \refeq(smallr-af) imply 
\[
|K^{(j,k)}_1(\rho)|\leq 
C_{jk}(\Mg \Hg(r^{2}M)(\rho)+ \Mg\Hg(\tilde{M})(\rho))
\]
where the last term is necessary only for $j=0$. Then, 
again by virtue of the weighted inequality and Hardy's 
inequality we obtain  
\begin{align}
& \left(\int_{|y|<1}\left|\frac{K^{(j,k)}_1(|y|)}
{|y|^{2}}\right|^p dy \right)^{1/p} 
\leq C \left(\int_{\R} 
(|M(r)r^2|^p + |\tilde{M}(r)|^p) r^{m-1-2p}d\rho \right)^{1/p} 
\notag \\
& \qquad \leq C \left(\int_{\R} |M(r)|^p r^{m-1}d\rho \right)^{1/p}
\leq C \|V\p \|_1 \|u\|^p.
\end{align}
Since $\frac{m}{m-2}<p'<\frac{m}{m-3}$, 
$|x|^{-(m-4)p'}$ is integrable on $|x|\leq 1$ and $p'\leq p$. 
It follows by Minkowski's inequality  
\[
\left\|\left(\int_{|y|<1}\left|\frac{V \f(x-y)}
{|y|^{m-4}}\right|^{p'}dy \right)^{1/p'}\right\|_p 
\leq C\|V\f\|_p 
\left(\int_{|y|<1}\frac{dy}{|y|^{p'(m-4)}}\right)^{1/p'}. 
\]
Thus, $\|I_1\|_p\leq C\|u\|_p$ as well. 
This completes the proof of the lemma.
\edpf 

\section{Proof of \refthb(theo) for even $m \geq 6$} 

We now prove \refth(theo) when the space dimension 
$m \geq 6$ is even, viz. we prove that $Z_{\log}$ and $Z_s$ 
defined respectively by \refeq(loge) and \refeq(e-zs) 
satisfy 
\bqn \lbeq(e-et)
\|Z_s u\|_p\leq C \|u\|_p, \quad 
\|Z_{\log} u\|_p\leq C \|u\|_p, \quad 1<p<m/2.  
\eqn 
We prove \refeq(e-et) for $Z_{s}$ first and comment 
on how to modify the argument for proving \refeq(e-et) 
for $Z_{\log}$ at the end of the section. We take 
the orthonormal basis $\{\f_1, \dots, \f_d\}$ of 
the $0$ eigenspace $\Eg$ of $H$ and represent      
$Z_{s}u(x) = i\pi^{-1}\sum_{l=1}^d {Z}_{s,l}u(x)$,  
where, for $l=1,\dots, d$,   
\bqn 
{Z}_{s,l}u  = 
\int_0^{\infty}G_0(\lam)|V\f_l\ra 
\la V\f_l| G_0(\lam)-G_0(-\lam))u \ra 
F(\lam)\frac{d\lam}{\lam}. \lbeq(Zn6) 
\eqn 
and we prove that ${Z}_{s,l}u$, 
$l=1,\dots, d$ satisfies \refeq(e-et) omitting the index 
$l$ in what follows. Recall $\n=(m-2)/2$. 

We want to apply the argument used for odd dimensions 
also to the even dimensional case as much as possible. For 
this purpose we represent ${Z}_{s}u(x)$ as in \refeq(Wsm-rep) 
below. We write $M(r)=M(r,V\f \ast \check{u})$ 
as previously and define $Q_{jk}^{a,b}(\rho)$ for 
$j, k=0, \dots, \n$ and $a,b>0$ by 
\bqn 
Q_{jk}^{a,b}(\rho) = \frac{(-1)^{j+1}|\Si|}{(1+2a)^{j+2}} 
\int_{0}^\infty \lam^{j+k-1} e^{i\lam(1+2b)\rho} 
\Fg (r^{j+1} M^a)(\lam) 
F(\lam) d\lam, \lbeq(kjkab) 
\eqn 
where $M^a(r)=M(r/(1+2a))$. When $j=0$,  
$Q_{0k}^{a,b}(\rho)$ may also be expressed as 
\bqn 
Q_{0k}^{a,b}(\rho) = \frac{i|\Si|}{(1+2a)^{2}} 
\int_{0}^\infty \lam^k e^{i\lam(1+2b)\rho} 
\Fg (\widetilde{M}^a)(\lam) 
F(\lam) d\lam \lbeq(k0kab) 
\eqn 
by using \refeq(usformula-add).      
Set $\Phi_{jk}(\lam)=\lam^{j+k-1} F(\lam)$. 
Then $\Phi_{jk}\in C_0^\infty(\R)$ if $j+k>0$ and, 
\begin{align}
Q_{jk}^{a,b}(\rho)& =(-1)^{j+1}2\pi(1+2a)^{-(j+2)} 
 \{(\Fg^\ast\Phi_{jk}) 
\ast \Hg(r^{j+1} M^a)\}((1+2b)\rho) \notag \\
& \absleq 
C(1+2a)^{-(j+2)}\{\Mg \Hg(r^{j+1} M^a)\}((1+2b)\rho). 
\end{align}

\bglm Let ${Q}_{jk}^{a,b}(\rho)$ be as in \refeq(kjkab) 
or \refeq(k0kab). Then,    
\bqn 
Z_{s}u(x) = \sum_{j,k=0}^\n 
T^{(a)}_j  T^{(b)}_k \left[ \int_{\R^m} 
\frac{(V\f)(x-y) Q^{a,b}_{jk}(|y|)}{|y|^{m-2-k}}dy 
\right]\equiv  
\sum_{j,k=0}^\n {Z}_{s}^{(j,k)}u(x) .
\lbeq(Wsm-rep) 
\eqn 
\edlm 
\bgpf In the right of \refeq(Zn6), use formula 
\refeq(usformula) for 
$\la V\f ,(G_0(\lam)-G_0(-\lam))u \ra$ and 
\refeq(co-kerb) for $G_0(\lam)$, viz. 
\bqn 
G_0(\lam)\p(x) = \sum_{k=0}^\n \lam^k 
T_k^{(b)}\left[ 
\int_{\R^m} 
\frac{e^{i\lam(1+2b)|y|}}
{|y|^{m-2-k}}
(V\f)(x-y)dy 
\right]. 
\eqn 
Then, we see that ${Z}_{s}u(x)$ is the integral 
with respect to $\lam\in (0,\infty)$ of $|\Si|$ 
times 
\[
\sum_{j,k=0}^\n 
T^{(a)}_j  T^{(b)}_k 
\left[
\frac{(-1)^{j+1}}{(1+2a)^{j+2}}
\lam^{j+k-1}
\left(\frac{e^{i\lam(1+2b)|y|}}
{|y|^{m-2-k}}\ast V\f \right)
\Fg (r^{j+1} M^a)(\lam)
\right]
F(\lam).
\] 
Integrating with respect to $\lam$ first via Fubini's  
theorem yields \refeq(Wsm-rep). 
\edpf 

\subsection{Estimate of $\|{Z}_{s}^{(j,k)}u\|_p$ 
for $1<p<\frac{m}{m-1}$} 

If $(j,k)\not=(\n,\n)$, the repetition of the argument 
modulo change of variables and 
the superposition via $T_j^{(a)}T^{(b)}_k$ 
of the previous section proves  
$\|Z_s^{(j,k)}u\|_p\leq C \|u\|_p$ for 
$1<p<\frac{m}{m-1}$. We first check this.   

\subsubsection{Estimate of $\|{Z}_{s}^{(j,k)}u\|_p$, 
$(j,k)\not=(\n,\n)$ for $1<p<\frac{m}{m-1}$. }

We use the following estimate which is basically 
the change of variable formula. Define 
\bqn \lbeq(n)
N_{\s}^{a,b}(u) = \left( \int^\infty_0 
|\Mg(r^{\s} M^{a})((1+2b)\rho )|^p 
\rho ^{m-1-p(m-1)} d\rho  \right)^{1/p}.
\eqn 
\bglm \lblm(njl) Let $0\leq  \s \leq m-1$ and 
$1<p<m/(m-1)$. Then, for any $m/(1+\s)\leq q\leq \infty$ 
we have 
\bqn \lbeq(es-1)
N_{\s}^{a,b}\leq C (1+2a)^{\frac{m}{p}-(m-1-\s)}
(1+2b)^{m-1-\frac{m}p}
(\|V\p\|_1+ \|V\p\|_{q})\|u\|_p.  
\eqn 
\edlm 
\bgpf 
By using changes of variables and weighted inequality 
we have 
\begin{multline*}
N_{\s}^{a,b} =(1+2b)^{m-1-\frac{m}p} 
\left( \int^\infty_0 
|\Mg(r^{\s} M^{a})(\rho )|^p 
\rho ^{m-1-p(m-1)} d\rho  \right)^{1/p}  \\
\leq  C (1+2a)^{\frac{m}{p}-(m-1-\s)}
(1+2b)^{m-1-\frac{m}p} \left( \int^\infty_0 
|M(r)|^p r^{m-1-p(m-1-\s)} dr \right)^{1/p}.
\end{multline*}
Denote $\kappa=m-1-\s$. Then $0\leq \kappa \leq m-1$ 
and the last integral is bounded 
by 
\begin{align*}
& \left( \int^\infty_0 
|M(r)|^p r^{m-1} dr \right)^{1/p}
+ \left( \int^1_0 |M(r)|^p 
r^{m-1-p\kappa} dr \right)^{1/p} \\
& \leq \|V\p \ast u\|_p+ 
\left\||x|^{-\kappa}(V\p\ast u)(x)
\right\|_{L^p(|x|<1)} \\ 
& \leq 
(\|V\p\|_1+ \||x|^{-\kappa}\|_{L^{\frac{m}{\kappa}}
(|x|\leq 1)}
\|V\p\|_{q}) \|u\|_p. 
\end{align*}
for any $q\in [m/(1+\s), \infty]$ 
by H\"older's and weak-Young's inequality. 
\edpf 

\bglm \lblm(zsq2) Suppose $1<p<\frac{m}{m-1}$. Then, 
for $2\leq j \leq \n$ and $0\leq k\leq \n$ such that 
$(j,k)\not=(\n, \n)$,  
\bqn \lbeq(zsq2)
\|{Z}_{s}^{(j,k)}u\|_p \leq C \|u\|_p, \quad u \in 
C_0^\infty(\R^m).
\eqn
\edlm 
\bgpf Minkowski's inequality and then Young's 
inequality yield   
\bqn 
\|{Z}_{s}^{(j,k)}u\|_p \leq 
\|V\f\|_1 \cdot T^{(a)}_j  T^{(b)}_k \left[
\left\| \frac{Q^{a,b}_{jk}(|x|)}{|x|^{m-2-k}}\right\|_p  
\right]. \lbeq(k-4) 
\eqn 
We apply integration by parts $k+1$ times  
to \refeq(kjkab) using that 
$e^{i\lam(1+2b)\rho }= (i(1+2b)\rho )^{-(k+1)}
\pa_{\lam}^{k+1} e^{i\lam(1+2b)\rho }$.
Then, \reflm(add-cut) and \reflm(max) imply 
\begin{align}\lbeq(devel)
& Q^{a,b}_{jk}(\rho )= \sum_{l=0}^{k+1} 
\frac{(-1)^{j+1}|\Si|}{(1+2a)^{j+2}} 
\left(\frac{1}{-i(1+2b)\rho }\right)^{k+1} 
\begin{pmatrix} k+1 \\ l \end{pmatrix} \notag \\
& \times \int_{0}^\infty 
e^{i\lam(1+2b)\rho } 
\left(\lam^{j+k-1}F(\lam)\right)^{(k+1-l)} 
\Fg ((-i)^l r^{j+l+1} M^{a})(\lam) d\lam \\
& \absleq \sum_{l=0}^{k+1} 
\frac{C}
{(1+2a)^{j+2}(1+2b)^{k+1}\rho ^{k+1}} 
\Mg(r^{j+l+1} M^{a})((1+2b)\rho ).  \lbeq(devel-a) 
\end{align}
Use \reflm(njl) with $s \equiv j+l+1\leq m-1$ for 
$(j,k)\not=(\n,\n)$ to obtain   
\bqn 
\left\|\frac{Q_{jk}^{a,b}(|x|)}{|x|^{m-2-k}}\right\|_p 
\leq C (1+2a)^{\frac{m}{p}-(m-k-1)}
(1+2b)^{m-2-\frac{m}p-k}\|u\|_p.  \lbeq(d-1)
\eqn   
We then plug this to \refeq(k-4) and use $m-k-1\geq j+2$. 
Then, 
\begin{align*}
& \|{Z}_{s2}^{(j,k)} u\|_p 
\leq C_{mjk} T_j^{(a)} T_k^{(b)}[
(1+2a)^{\frac{m}{p}-(j+2)}
(1+2b)^{m-2-\frac{m}p-k}]\|u\|_p \\
& \leq C \|u\|_p 
\left(\int_0^\infty 
\frac{(1+2a)^{\frac{m}{p}-(j+2)}}{(1+a)^{(2\n-j+\frac12)}}
\frac{da}{\sqrt{a}}\right) 
\left(\int_0^\infty 
\frac{(1+2b)^{m-2-\frac{m}p-k}}
{(1+b)^{(2\n-k+\frac12)}}
\frac{db}{\sqrt{b}}\right) .
\end{align*} 
Since $\frac{m}{p}-(j+2)-2\n +j= \frac{m}{p}-m<0$ and 
$m-2-\frac{m}p<-1$ for $1<p<\frac{m}{m-1}$, 
the integrals on the right are finite and the lemma 
follows. 
\edpf  

We can repeat the argument used for odd dimensional 
$m \geq 5$ case also  for 
\bqn \lbeq(sum-jk)
Z_{s2}^{(0,k)}u+ Z_{s2}^{(1,k)}u =
\int_{\R^m} \frac{(V\f)(x-y)}{|y|^{m-2-k}}
T_k^{(b)}(T_0^{(a)}Q_{0k}^{(a,b)}(|y|)+ 
T_1^{(a)}Q_{0k}^{(a,b)}(|y|))dy 
\eqn 
and obtain the following lemma. 

\bglm \lblm(zsq3) For $1<p<\frac{m}{m-1}$, 
there exists a constant $C>0$ such that 
\bqn 
\|({Z}_{s}^{(0,k)}+ {Z}_{s}^{(1,k)})u\|_p 
\leq C \|u\|_p, \quad k=0,\dots, \n.
\eqn
\edlm 
\bgpf We apply integration by parts $k+1$ times 
to \refeq(k0kab) and \refeq(kjkab) as in the $j\geq 2$ 
cases. This respectively produces the following: 
\begin{align}
Q_{0k}^{a,b}(\rho )
& = \frac{-i^{k}k!(\Fg \widetilde{M^a})(0)|\Si|}
{(1+2a)^2(1+2b)^{k+1}\rho ^{k+1}}
- i^k |\Si|
\sum_{l=0}^{k+1}C_{k+1,l} Q_{0k,l}^{a,b}(\rho )  \lbeq(b+1) \\
Q_{1k}^{a,b}(\rho )& = 
\frac{i^{k+1}k!\Fg(r^2 M^a)(0)|\Si|}
{(1+2a)^3(1+2b)^{k+1}\rho ^{k+1}}
+i^{k+1}|\Si|\sum_{l=0}^{k+1} 
C_{k+1,l} Q_{1k,l}^{a,b}(\rho ), \lbeq(b+2) 
\end{align}
where the integral terms $Q_{0k,l}^{a,b}(\rho )$ and 
$Q_{1k,l}^{a,b}(\rho )$ are given 
\begin{align}
Q_{0k,l}^{a,b}(\rho ) & = \int_0^\infty 
\frac{e^{i\lam(1+2b)\rho }(\lam^k F(\lam))^{(k+1-l)}
(\Fg ((-ir)^l \widetilde{M^a})(\lam))}
{(1+2a)^2(1+2b)^{k+1}\rho ^{k+1}}d\lam \notag \\ 
& \absleq C 
\frac{\Mg (r^l \widetilde{M^a})((1+2b)\rho))}
{(1+2a)^2(1+2b)^{k+1}\rho ^{k+1}}.  \lbeq(fg-1) \\
Q_{1k,l}^{a,b}(\rho )& = (-i)^{l} 
\int_0^\infty \frac{e^{i\lam(1+2b)\rho }(\lam^k F(\lam))^{(k+1-l)}
\Fg(r^{2+l} M^a)(\lam))}
{(1+2a)^3(1+2b)^{k+1}\rho ^{k+1}}
d\lam \notag \\
& \absleq C \frac{\Mg(r^{2+l} M^a)((1+2b)\rho)}
{(1+2a)^3(1+2b)^{k+1}\rho ^{k+1}}.
\lbeq(fg-2) 
\end{align} 
We have  
$\Fg (\widetilde{M^a})(0)
= \Fg (r^2 M^a)(0)
= (1+2a)^3 \int_0^\infty r^2 M(r) dr$ 
by virtue of \refeq(MtildeM-1) and, 
elementary computations show   
\[
T_1^{(a)} [i]= T_0^{(a)}[(1+2a)]
= 2\n\pi^{-\frac12}C_m^{-1}\Ga(1/2)\Ga(2\n-1). 
\]
It follows that the sum of the superposition via 
$T_0^{(a)}$ of the boundary term of \refeq(b+1) 
and that via $T_1^{(a)}$ of the one of \refeq(b+2) vanishes:  
\bqn 
\frac{i^k k!}
{(1+2b)^{k+1}\rho^{k+1}}
\left(\int_0^\infty r^2 M(r) dr \right)
(T_1^{(a)}[i] -T_0^{(a)}[(1+2a)])=0.  
\lbeq(vabo)
\eqn 
For $1<p<\frac{m}{m-1}$, $\rho^{m-1-p(m-1)}$ is an 
$A_p$ weight on $\R$ and we have the identity: 
\bqn \lbeq(Ha-def)
\widetilde{M^a}(r) = \int_r^\infty s M^a (s) ds 
= (1+2a)^2 \tilde{M}((1+2a)^{-1}r). 
\eqn 
It follows by using \reflm(ap) and 
the change of variable that  
\begin{align}
& \left\|\frac{Q_{0k,l}^{a,b}(|x|)}{|x|^{m-k-2}}\right\|_p 
\leq 
\frac{C(1+2b)^{m-1-\frac{m}{p}}}{(1+2a)^2(1+2b)^{k+1}} 
\left( \int^\infty_0 
|r^{l} \widetilde{M^{a}}(r)|^p 
r^{m-1-p(m-1)} dr \right)^{1/p} \notag \\
& = C \frac{(1+2a)^{\frac{m}{p}-(m-1-l)}}
{(1+2b)^{\frac{m}{p}-(m-k-2)}} \left( \int^\infty_0 
|\tilde{M}(r)|^p r^{m-1-p(m-1-l)} dr \right)^{1/p}. 
\lbeq(last-1)
\end{align}
Since $m-p(m-1-l)>0$, Hardy's inequality applies and, 
because $0 \leq p(m-4-k)\leq p(m-3-l)<m$ for $m \geq 6$,  
we have as previously    
\begin{align}
& \refeq(last-1) \leq \frac{C(1+2a)^{\frac{m}{p}-(m-1-l)}}
{(1+2b)^{\frac{m}{p}-(m-k-2)}}
\left( \int^\infty_0 
|M(r)|^p r^{m-1-p(m-3-l)} dr \right)^{1/p}  \notag \\
& \leq \frac{C(1+2a)^{\frac{m}{p}-(m-k-2)}}
{(1+2b)^{\frac{m}{p}-(m-k-2)}}(\|V\p\|_1+ \|V\p\|_{m/2})
\|u\|_p, \ 0\leq l\leq k+1. \lbeq(d6-0)
\end{align} 
Since $\frac{m}{p}-(m-k-2)-(2\n+1)< -1 $ 
and $\frac{m}{p}-(m-k-2)+ 2\n-k+1 >2$, 
$(1+2a)^{\frac{m}{p}-(m-k-2)}
(1+a)^{-(2\n+\frac12)}a^{-\frac12}$ 
and 
$(1+2b)^{-\frac{m}{p}+(m-k-2)}
(1+b)^{-(2\n-k+\frac12)}b^{-\frac12}$ 
are both integrable over $(0,\infty)$.
Thus, \refeq(d6-0) implies 
\bqn 
T_0^{(a)}T_k^{(b)} \left[ 
\left\|
\frac{Q_{0k,l}^{a,b}(|x|)}{|x|^{m-k-2}}
\right\|_p 
\right] \leq C \|u\|_p.  \quad \ 0\leq l\leq k+1 \lbeq(g-1).
\eqn 
Entirely similarly, starting from \refeq(fg-2), we obtain  
\begin{align}
& \left\|
\frac{Q_{1k,l}^{a,b}(|x|)}
{|x|^{m-k-2}}
\right\|_p 
\leq \frac{
C(1+2b)^{m-1-\frac{m}{p}}
}
{(1+2a)^3(1+2b)^{k+1}} 
\left( 
\int^\infty_0 
|r^{2+l}M^a(r)|^p 
r^{m-1-p(m-1)} dr 
\right)^{1/p}  \notag \\
& 
\leq \frac{
C(1+2a)^{\frac{m}{p}-(m-l)}
}
{
(1+2b)^{\frac{m}p-(m-k-2)}
} 
\left( 
\int^\infty_0 
|M(r)|^p 
r^{m-1-p(m-3-l)} dr 
\right)^{1/p} \notag \\
& \leq 
\frac{
C(1+2a)^{\frac{m}{p}-(m-k-1)}
}
{
(1+2b)^{\frac{m}p-(m-k-2)}
} (\|V\p\|_1+ \|V\p\|_{\frac{m}{3}})
\|u\|_p, \ 0\leq l\leq k+1. \lbeq(d7-m)
\end{align}
The extra decaying factor $(1+2a)^{-1}$ of \refeq(d7-m) 
compared to \refeq(d6-0) cancels the extra increasing 
factor $(1+a)$ of $T_1^{(a)}$ compared to $T_0^{(a)}$ and 
we have also   
\bqn 
T_1^{(a)}T_k^{(b)} 
\left[ 
\left\|
\frac{Q_{0k,l}^{a,b}(|x|)}{|x|^{m-k-2}}
\right\|_p 
\right] \leq C \|u\|_p, \quad 0\leq l \leq k+1.
\lbeq(g-2)
\eqn 
Estimates \refeq(g-1) and \refeq(g-2) implies the lemma 
by virtue of \refeq(sum-jk), \refeq(b+1), \refeq(b+2) 
and Young's inequality as in \refeq(k-4). 
\edpf 

\subsubsection{Estimate of $Z_{s}^{(\n,\n)}$ for $1<p<m/(m-1)$} 

In this subsection, we prove 
\bqn 
\|Z_{s}^{(\n,\n)} u\|_p \leq C \|u\|_p, \quad 1<p<m/(m-1).
\eqn 
If we apply the method of previous section for proving 
this, then, in the integration by parts formula \refeq(devel), 
the power $j+l+1$ of $\Fg((-i)^l r^{j+l+1})$ can reach   
$m$, which is too large to be controlled by using the weighted 
inequality and we have to use the different method. 
We first pinpoint the problem by applying the previous 
argument. 

We apply integration by parts $\n-1$ times to \refeq(kjkab): 
\begin{multline} 
Q_{\n\n}^{a,b}(\rho)= \sum_{l=0}^{\n-1} 
\frac{i^{\n-1-l}(-1)^{\n-1} C_{\n-1,l}|\Si|}
{(1+2a)^{\n+2}(1+2b)^{\n-1}\rho^{\n-1}}  \\
 \times \int_0^\infty 
e^{i(1+2b)\rho\lam}(\lam^{2\n-1} F)^{(\n-1-l)}
\Fg(r^{\n+l+1}M^{a})(\lam)d\lam 
\equiv \sum_{l=0}^{\n-1} (L^{a,b}_l u)(\rho), \lbeq(new)
\end{multline}
where the definition should be obvious.  
For $L_l^{a,b} u$ with $0\leq l\leq \n-2$, 
we further apply integration by parts twice to see     
\begin{align} 
L^{a,b}_l u(\rho) & = \sum_{j=0}^2 
C_{lj}\frac{\{\Fg^\ast {\Phi}_{\n,j}\ast 
\Hg(r^{\n+l+1+j}M^{a})\}(\rho)}
{(1+2a)^{\n+2}(1+2b)^{\n+1}\rho^{\n+1}} \notag \\
& \absleq C \sum_{j=0}^2 
C_{lj}\frac{\Mg(r^{\n+l+1+j}M^{a})(\rho)}
{(1+2a)^{\n+2}(1+2b)^{\n+1}\rho^{\n+1}}
\lbeq(Ll)
\end{align}
Here $\n+l+j+1 \leq 2\n+1\leq m-1$. It follows by virtue of 
\reflm(njl) that   
\[
\left\|\frac{L^{a,b}_l}{\rho^{m-2-\n}}\right\|_p 
\leq C_\ep \frac{(1+2a)^{\frac{m}{p}}(1+2b)^{m-1-\frac{m}{p}}}
{(1+2a)^{\n+2}(1+2b)^{\n+1}}
(\|V\p\|_{1}+ \|V\p\|_{\frac{m}{\n+2}}\|u\|_p, 
\]
and, as $\frac{m}{p}-(\n+2)-\n = 
m\left(\frac1{p}-1\right) <0$ 
and 
$(m-1-\frac{m}{p})-(\n+1)-\n = -\frac{m}{p}<0$, 
\bqn \lbeq(Ll3)
\left\|T^{(a)}_\n T^{(b)}_\n 
\left[ \frac{L_l}{\rho^{m-2-\n}} \right] \right\|_p 
\leq C(\|V\p\|_{\frac{m}{\n+2}}+ \|V\p\|_{1})\|u\|_p, 
\  0\leq l\leq \n-2.
\eqn 
Thus, we have only to prove for $1<p<m/(m-1)$ that    
\bqn 
\left\|T^{(a)}_\n T^{(b)}_\n \left[\frac{L_{\n-1}^{a,b}(\rho)}
{\rho^{m-2-\n}}\right]\right\| \leq C \|u\|_p.  
\eqn 
The problem here is that, if we apply integration by parts 
twice to obtain the factor $\rho^{-(\n+1)}$ as in \refeq(Ll), 
we have $\Fg(r^{2\n+2} M)$ there and $2\n+2=m$ is too big 
to be controlled. Thus, we integrate with respect to $a, b$ 
first: 
\begin{align}
& T_\n^{(a)} T_\n^{(b)} 
\left(
\frac{L_{\n-1}^{a,b}(\rho)}{\rho^{m-\n-2}}
\right) 
= \frac{C}{\rho^{m-3}} 
\int_0^\infty 
\left\{
e^{i\lam\rho} \lam^{m-3}
\left(
\int_0^\infty \frac{(1+2b)^{1-\n}}{(1+b)^{\n+\frac12}}
e^{2i\lam\rho {b}}\frac{db}{\sqrt{b}}
\right) 
\right.  \notag \\
& \times \left.
\int_{\R}e^{-i\lam{r}}
\left(
\int_0^\infty \frac{(1+2a)^{\n-1}}{(1+a)^{\n+\frac12}}
e^{-2ia\lam{r}}
\frac{da}{\sqrt{a}}
\right) \lbeq(m-1)
r^{m-2}M(r)dr 
\right\}
F(\lam)d\lam.  
\end{align}
Define, for $t>0$,  
\begin{gather}
g(t) = \int_0^\infty 
\left(1+ \frac{b}{2t}\right)^{-\n-\frac12} 
\left(1+ \frac{b}{t}\right)^{-\n+1} 
e^{ib}\frac{db}{\sqrt{b}}, \lbeq(f-2b) \\  
h_\pm (t) = \int_0^\infty 
\left(1+ \frac{a}{2t}\right)^{-\n-\frac{1}2} 
\left(1+ \frac{a}{t}\right)^{\n-1} 
e^{\pm ia}\frac{da}{\sqrt{a}}    \lbeq(f-2a)
\end{gather}
so that for $\lam>0$
\bqn 
\int_0^\infty \frac{(1+2a)^{\n-1}}{(1+a)^{\n+\frac12}}
e^{\pm 2ia\lam{r}}
\frac{da}{\sqrt{a}}= 
\frac1{\sqrt{2\lam\rho}} h_{\pm}(\lam\rho), 
\eqn 
and the like for $g(\lam\m)$. Then, with 
$\chi_\pm(r)=1$ being the characteristic function of 
$\pm r>0$,   
\begin{gather} \lbeq(f-3)
T_\n^a T_\n^b 
\left(\frac{L_{\n-1}^{a,b}(\rho)}{\rho^{m-2-\n}}\right) 
= \frac{C(m)}{2\rho^{m-\frac52}}
\int_{\R} (L_{+}(\rho,r)+ L_{-}(\rho,r))
r^{m-2}M(r)dr, \\ 
L_{\pm}(\rho,r)= \frac{\chi_\pm(r)}{|r|^\frac12}
\int_0^\infty e^{i\lam(\rho-r)}
\lam^{m-4}g(\lam\rho)h_{\mp}(|r|\lam)F(\lam) d\lam.  
\lbeq(f-1)
\end{gather} 
We have the following lemma. 

\bglm \lblm(gpm) Suppose that $f \in C^\infty([0,\infty))$ 
satisfies 
\bqn  
|f^{(j)}(b)| \leq C_jb^{-(j+1)} \quad j=0,1, \dots 
\lbeq(a1) 
\eqn 
Let 
\[
h_\pm(t) = \int_0^\infty e^{\pm ib}f(b/t)\frac{db}{\sqrt{b}}.
\]
Then, $h_{\pm}(t)$ is $C^\infty$ for $t>0$ and 
satisfies the following properties. 
\ben 
\item[{\rm (1)}] $\tilde{h}_{\pm}(x)=h_{\pm}(1/x)$ is of class 
$C^\infty([0,1])$, hence the limit 
$\lim_{t\to \infty} h_{\pm}(t)= C_m$ exists 
and for $t\geq 1$, 
$|h_\pm ^{(j)}(t)|\leq C_j t^{-j-1}$, $j=1,2,\dots$. 
\item[{\rm (2)}] For $0<t<1$, 
$|t^j h_{\pm}^{(j)} (t)|\leq C_j \sqrt{t} \leq C_j$, 
$j=0,1, \dots$.
\een  
\edlm 
\bgpf We consider $h_{+}(t)$ only and omit the $+$-sign. 
The proof for $h_{-}(t)$ is similar. 
It is easy to see by differentiating under the sign of 
integration that $h(t)$ is $C^\infty$ for $t>0$. 
We first consider the case $t>1$. Splitting the interval 
of the integral, we write  
\[
h(t)=\left(\int_0^1 + \int_1^\infty\right) 
f\left(\frac{b}{t}\right)e^{ib}\frac{db}{\sqrt{b}}
\equiv {h}_{1}(t)+ {h}_{2}(t) 
\]
and $\tilde{h}_j(x)= h_j(1/x)$, $j=0,1$. 
It is obvious that $\tilde{h}_{1}(x)$ is $C^\infty$ 
up to $x=0$. To see the same for $\tilde{h}_{2}(x)$,  
we perform integration by parts $n$ times. Then, 
\begin{align}
& i^n \tilde{h}_{2}(x)= \int_1^\infty 
f(bx)b^{-\frac12} (e^{ib})^{(n)}db \notag \\
& = \sum_{j=0}^{n-1} (-1)^{j+1}
\pa_b^j\left(\frac{f(bx)}{\sqrt{b}}\right) \left. 
(e^{ib})^{(n-j-1)}\right\vert_{b=1} 
+(-1)^n \int_1^\infty \pa_b^n \left(\frac{f(bx)}{\sqrt{b}}
\right) e^{ib}db. \lbeq(o-1)
\end{align}
Here, Leibniz's formula implies  
\[
\pa_b^n \left(\frac{f(bx)}{\sqrt{b}}\right) 
=\sum_{j=0}^n C_{nj}
f^{(j)}(bx)(bx)^j b^{-\frac12-n} 
\]
and the boundary term in \refeq(o-1)
\[
\sum_{j=0}^{n-1} (-1)^{j+1}
\pa_b^j\left(\frac{f(bx)}{\sqrt{b}}\right)\left.  
(e^{ib})^{(n-j-1)}\right\vert_{b=1} 
=\sum_{j=0}^n C_{nj} f^{(j)}(x)x^{j}
\]
is evidently $C^\infty$. Since $\pa_y^{k} (f^{(j)}(y)y ^j)$ 
is bounded for any $j,k=0,1, \dots$ and 
\[
\pa_x^k \left(\sum_{j=0}^n C_{nj}
f^{(j)}(bx)(bx)^j b^{-\frac12-n} \right) 
= \sum_{j=0}^n C_{nj}
\left. \pa_y^{k} (f^{(j)}(y)y ^j)\right\vert_{y=bx} 
b^{-\frac12-n+k}, 
\]
the integral in the right of \refeq(o-1) produces 
a function which is of class $C^{n-1}([0,1])$. Since 
$n$ is arbitrary, this proves (1). For proving (2), 
notice that if $\alpha(t)$ and $\beta(t)$ satisfy 
\bqn 
|t^j \alpha^{(j)}(t)|\leq C_j \sqrt{t}, \quad 
|t^j \beta^{(j)}(t)|\leq C_j, \quad j=0,1, \dots \lbeq(o-2)
\eqn 
for $0<t<1$, then, by Leibniz' formula 
$\gamma(t)=\alpha(t)\beta(t)$ satisfies 
\bqn \lbeq(n-t)
|t^j \gamma^{(j)}(t)|\leq C_j \sqrt{t}, \quad 0<t<1, \quad 
j=0,1, \dots.
\eqn 
It is obvious that $\sqrt{t}$ satisfies \refeq(n-t) 
and so does $h_{1}$ as 
\bqn 
h_{1}(t) = \sqrt{t} \int^1_0 e^{itb}f(b)\frac{db}{\sqrt{b}}
\eqn  
is an entire function times $\sqrt{t}$.  
\bqn 
h_{2}(t) = \sqrt{t} 
\int^\infty_1 e^{itb}f(b) 
\frac{db}{\sqrt{b}} \equiv \sqrt{t} \ka(t). 
\eqn 
We show for $0<t<1$ that 
$|t^n \ka^{(n)}(t)| \leq C_n$, $n=1,2, \dots$.
This will follow from  
\bqn \lbeq(o-5) 
|(t^n \ka)^{(n)}(t)|\leq C_n, \quad 0<t<1, \quad n=0,1, \dots 
\eqn
by virtue of Leibniz' formula
$(t^n \ka)^{(n)}= \sum_{k=0}^n C_{nk} t^{n-k}\ka^{(n-k)}$. 
By integration by parts we have 
\begin{align*}
&(it)^n \ka(t)= \int_1^\infty (\pa_b^n e^{itb})f(b)  
\frac{db}{\sqrt{b}} \\
& = \sum_{j=0}^{n-1}(-1)^{j+1}
\pa_b^j \left(\frac{f(b)}{\sqrt{b}}\right)
\left. \pa_b^{n-j-1}(e^{itb})\right\vert_{b=1} + 
\int_1^\infty e^{itb}
(f(b)b^{-\frac12})^{(n)}db.
\end{align*}
The boundary term is a polynomial of $t$ and the integral is 
$n$ times continuously differentiable and a fortiori 
$(t^n \ka(t))^{(n)}\leq C$ for $0<t<1$. 
\edpf 

\bglm \lblm(p-1) 
Suppose that $g(t)$ and $h_\pm(t)$ are functions of 
$t>0$ of class $C^\infty$ and they satisfy following 
properties replacing $f$: 
\ben 
\item[{\rm (a)}] The limit $\lim_{t\to \infty} f(t)$ exists. 
\item[{\rm (b)}]  
$ |t^j f^{(j)}(t)|\leq C_j \left\{
\br{lll} t^{-1} ,  \quad &  1< t, & j=1,2,\dots, \\ 
\sqrt{t}, \quad  & 0<t<1, & j=0,1, \dots.
\er \right.$. 
\een   
Let $\s\geq 0$ be an integer and let $\tilde{L}_\pm(\rho,r)$ be 
defined by 
\bqn \lbeq(dlt)
\tilde{L}_{\pm}(\rho,r)= \chi_\pm(r)
\int_0^\infty e^{i\lam(\rho-r)}
\lam^{\s}g(\lam\rho)h_{\mp}(\pm r\lam)F(\lam) d\lam  
\eqn 
where $F\in C^\infty_0(\R)$ be such that $F(\lam)=1$ 
in a neighborhood of $\lam=0$. Then, $\tilde{L}_\pm$ is $C^\infty$ 
with respect to $\rho,r>0$ and, for a constant $C>0$,  
\bqn \lbeq(dlt-con)
|\tilde{L}_\pm (\rho, r)| \leq C \la \rho-r  \ra^{-(\s+1)}
\eqn 
\edlm 
\bgpf We prove the lemma for $\tilde{L}_{+}$. The proof for 
$\tilde{L}_{-}$ is similar. It is obvious that $\tilde{L}_{+}(\rho,r)$ 
is bounded and we assume $|\rho-r|\geq 1$.  
We apply integration by parts $\s+1$ times to   
\[
\tilde{L}(\rho,r)=\frac{(-i)^{\s+1}}{(\rho-r)^{\s+1}} \int_0^\infty 
\left(\pa_\lam ^{\s+1}
e^{i\lam(\rho-r)}\right)
\lam^{\s}g(\lam\rho)h_{-}(r\lam)F(\lam) d\lam.
\]
By Leibniz' rule the derivatives of 
$\lam^{\s}g(\lam\rho)h_{-}(r\lam)F(\lam)$ 
of order $\ka$ is a linear combination of 
\bqn \lbeq(derva)
\lam^{\s-\alpha-\beta-\gamma}
(\lam\rho)^\beta g^{(\beta)}(\lam\rho)
(r^\gamma\lam) h_{-}^{(\gamma)}(r\lam) F^{(\delta)}(\lam)
\eqn 
over $(\alpha,\beta,\gamma,\delta)$ such that 
$\alpha+\beta+\gamma+\delta=\ka $ and $\alpha \leq \s$. 
This converges to $0$ as $\lam \to 0$ if $\ka\leq \s$. 
It follows that no boundary terms appear and 
$(\rho-r)^{\s+1}\tilde{L}(\rho,r)$ is a linear combination 
over $(\alpha,\beta,\gamma,\delta)$ such that 
$\alpha+\beta+\gamma+\delta=\s+1$ and   
$\alpha \leq \s$ of 
\[
I_{\alpha\beta\gamma\delta}(\rho,r)=\int_0^\infty 
e^{i(\rho-r)\lam} 
\lam^{\s-\alpha}\rho^\beta g^{(\beta)}(\lam\rho)
r^\gamma h_{-}^{(\gamma)}(r\lam) F^{(\delta)}(\lam)d\lam. 
\]
It suffices to show that $I_{\alpha\beta\gamma\delta}(\rho,r)$ is 
bounded. If $\delta\not=0$, then $c_0<\lam<c_1$ 
for constants $0<c_0<c_1<\infty$ when   
$F^{(\delta)}(\lam)\not= 0$, and \reflm(gpm) implies 
$|(\lam\rho)^\beta g^{(\beta)}(\lam\rho)
(r\lam)^\gamma h_{-}^{(\gamma)}(r\lam) F^{(\delta)}(\lam)|\leq C$ 
and 
\[
I_{\alpha\beta\gamma\delta}(\rho,r) \absleq C \int_{c_0}^{c_1}  
\lam^{\s-\alpha-\beta-\gamma} d\lam \leq C, \quad \mbox{if $\delta\not=0$}. 
\]
Thus, we assume $\delta=0$ in what follows. We may also assume 
$0<r<\rho<\infty$ by symmetry. We split the region of 
integration into three intervals: 
\[
(0,\infty) = (0,1/\rho) \cup [1/\rho,1/r] \cup (1/r,\infty) 
\]
and denote the integral over these intervals by 
$I_1$, $I_2$ and $I_3$ in this order so that 
$I_{\alpha\beta\gamma\delta}(\rho,r)= I_1+ I_2+ I_3$. \\
(1) If $0<\lam<1/\rho$, we have $0<r\lam<\rho\lam <1$ and 
\begin{gather*} 
(\rho\lam)^\beta g^{(\beta)}(\rho\lam) 
\absleq C \sqrt{\rho\lam}, \quad 
(r\lam)^\gamma h_{-}^{(\gamma)}(r\lam) 
\absleq C \sqrt{r\lam}
\end{gather*} 
It follows that 
\bqn \lbeq(I-1)
I_1\absleq 
C\int_0^{1/\rho} 
\lam^{\s-\alpha-\beta-\gamma}\sqrt{\rho{r}}\lam d\lam 
= C \sqrt{\frac{r}{\rho}}\leq C 
\eqn 
(2) If $1/\rho\leq  \lam \leq 1/r$, we have 
$0<r\lam <1< \rho\lam $ and we estimate as   
\begin{align*}
& (a)\ g(\rho\lam)|\absleq C, \ \  
(r\lam)^\gamma h_{-}^{(\gamma)}(r\lam)\absleq C 
\sqrt{\lam r} & \mbox{if $\beta=0$,\ $\gamma\not=0$}. \\
& (b)\ (\rho\lam)^\beta g^{(\beta)}(\rho\lam) 
\leq  C, \ \ h_{-}(r\lam)\absleq C\sqrt{r\lam}, 
& \mbox{if $\beta\not =0$,\ $\gamma=0$}. \\
& (c)\ (\rho\lam)^\beta g^{(\beta)}(\rho\lam) \absleq C, \ \ 
(r\lam)^\gamma h_{-}^{(\gamma)}(r\lam)
\absleq C\sqrt{r\lam}, \  & \mbox{if $\beta\not= 0$,\  
$\gamma\not=0$}. 
\end{align*}
In all cases, we have $\s-\alpha-\beta-\gamma= -1$ and 
\[
\lam^{\s-\alpha-\beta-\gamma}
(\lam\rho)^\beta g^{(\beta)}(\lam\rho)
(r^\gamma\lam) h_{-}^{(\gamma)}(r\lam) 
\absleq \lam^{-\frac12}\sqrt{r}. 
\]
It follows that 
\bqn \lbeq(I-2)
I_2 \absleq 
C \int_{1/\rho}^{1/r} \lam^{-\frac12}\sqrt{r}d\lam 
= 2C \sqrt{r}\left(\frac1{\sqrt{r}}-\frac1{\sqrt{\rho}}\right) 
\leq 2C. 
\eqn 
(3) Finally if $1<r\lam<\rho\lam$, then we likewise estimate 
\[
\lam^{-1}|(\lam\rho)^\beta g^{(\beta)}(\lam\rho)
(r^\gamma\lam) h_{-}^{(\gamma)}(r\lam)| 
C \left\{\br{ll} \lam^{-1}(r\lam)^{-1}, \ & 
\mbox{if $\beta=0, \ \gamma\not=0$} \\
\lam^{-1}(\rho\lam)^{-1}, \ &
\mbox{if $\beta\not=0, \ \gamma=0$}, \\
\lam^{-1}(\rho\lam)^{-1}(r\lam)^{-1}, \ & 
\mbox{if $\beta, \gamma\not=0$}. \er 
\right. 
\]
The right  hand side is all bounded by $C r^{-1}\lam^{-2}$ 
and 
\[
I_3 \absleq C \int_{1/r}^\infty \lam^{-2} r^{-1} d\lam 
= C.    
\]
This completes the proof.  
\edpf 

Now we can prove the following proposition. 
\begin{proposition} \lbprp(last-prop)
Let $m\geq 6$. Then, for 
$1\leq p <m/(m-1)$, we have  
\bqn \lbeq(res-znn)
\|Z_{s}^{(\n,\n)} u\|_p \leq C_p \|u\|_p .
\eqn 
\end{proposition} 
\bgpf We write $M(r)$ for $M_{V\f\ast \check u}$ as 
previously. 
By virtue of \refeq(f-3) and \refeq(Wsm-rep), 
we see that $\|Z_{s}^{\n\n} u\|_p$ is bounded by a 
constant times 
\[
\sum_{\pm}C\|V\f\|_1 \left(\int_0^\infty \rho^{m-1}
\left(\frac{1}{\rho^{m-\frac52}} \int_{\R}
|L_{\pm}(\rho,r)r^{m-2}M(r)| dr \right)^p d\rho 
\right)^{\frac1{p}}
\]
and \reflm(p-1) implies that the sum of is bounded by 
\bqn \lbeq(g-13) 
C\left( \int_0^\infty\left( 
\int_{\R} \rho^{\frac{m-1}{p}-m+\frac52} 
\la \rho-r \ra^{-(m-3)}|r^{m-\frac52}M(r)| dr\right)^p 
\rho \right)^{1/p}  
\eqn 
Define $\ka\equiv \frac{m-1}{p}-m+\frac52$. Then, 
$0 < \ka<3/2$ for $1< p <2(m-1)/(2m-5)$ and  
\begin{align}
& \rho^{\ka}\la \rho-r \ra^{-(m-3)}|r|^{m-\frac52} 
\leq C\la \rho-r\ra^{-(m-3-\ka)} \la r \ra^{\ka}
|r|^{m-\frac52} \notag \\
& \hspace{1.5cm} \leq C 
\left\{
\br{l} 
\la \rho-r\ra^{-(m-3-\ka)} |r|^{m-\frac52}
\quad \mbox{for}\ |r|\leq 1 \\
\la \rho-r\ra^{-(m-3-\ka)} |r^{\frac{m-1}{p}}|
\quad \mbox{for}\ |r|\geq 1. 
\er 
\right. \lbeq(g-13a)
\end{align}
Here $m-3-\ka > 1$ and $\la r \ra^{-(m-3-\ka)}$ 
is integrable on $\R$ if $m \geq 6$. 
Thus, \reflm(njl) implies that 
\begin{multline}
\refeq(g-13a) \leq C 
\left(\int_{|r|<1}
|r^{m-\frac52}M(r)|^p dr\right)^{\frac{1}{p}} + 
C \left(
\int_{|r|\geq 1}
 |M(r)|^p |r|^{m-1} dr \right)^{\frac{1}{p}} \\
\leq C (\|\p\|_{\s}+ \|\p\|_1) \|u\|_p, \lbeq(gfnal) 
\end{multline}
for any $\s>m-\frac52$.  
This completes the proof. Actually, the proof shows that  
\refeq(g-13) holds for $p$ such that $1<p<2(m-1)/(2m-5)$. 
\edpf

\subsection{Estimate of $\|Z_{s}u\|_p$ for $m/3<p<m/2$}

We prove in this section that $\|Z_s u\|_p\leq C \|u\|_p$ 
for $p$ such that $m/3<p<m/2$ for even $m \geq 6$.  
It suffices to show that $Z_s^{(j,k)}$ satisfies \refeq(zsq2) 
for $0\leq j,k\leq \n$. We prove this by virtually repeating 
the argument used for proving \reflm(lastlm-odd) for 
odd dimensions modulo the superpositions. 
Let $j \geq 1$ first. Then, we split $Z_s^{(j,k)}$ as in 
\refeq(split-odd) as 
\bqn \lbeq(split-even) 
Z_s^{(j,k)}u(x)= T_j^{(a)}T_k^{(b)} \left[ 
\left(\int_{|y|<1}+ \int_{|y|>1}\right) 
\frac{(V\f)(x-y) Q_{jk}^{a,b}(|y|)}{|y|^{m-2-k}}dy \right]
\eqn
and estimate the integrals over $|y|\leq 1$ and $|y|>1$ 
separately. We write $Q_{jk}^{a,b}(\rho)$ 
as follows as in \refeq(odd-ip) and apply integration by 
parts in a way similar to the one used in \refeq(smallr) 
to see that   
\begin{align}
& Q_{jk}^{a,b}(\rho) = |\Si| \int_{0}^\infty 
\frac{\lam^{j+k-1}F(\lam) e^{i\lam(1+2b)\rho} 
\pa_\lam^{j-1} \{\Fg (r^{2} M^a)(\lam)\}}
{(-i)^{j+1}(1+2a)^{j+2}} 
d\lam  \lbeq(kjkab-sub-1) \\
& = \sum_{l=0}^{j-1}C_{jkl}
\frac{((1+2b)\rho)^{l}}{(1+2a)^{j+2}} 
\int_{0}^\infty e^{i\lam(1+2b)\rho} \Phi_{jkl}
\Fg (r^{2} M^a)(\lam) d\lam \notag \\
& \absleq 
\frac{C(1+2b)^{j-1}}{(1+2a)^{j+2}} 
\Mg \Hg(r^{2} M^a)((1+2b)\rho)
\times \left\{ \br{cl} \rho^{j-1}, & \ \rho \geq 1, \\
1,  & \ \rho<1,  \er \right. 
\lbeq(kjkab-sub) 
\end{align} 
where $\Phi_{jkl}= (\lam^{j+k-1}F)^{(l)}$. 
We change variable and use \reflm(ap) for 
the $A_p$ weight $\rho^{m-2p}$ for $m/3<p<m/2$ 
to obtain 
\begin{align}
& \left(\int_0^\infty 
|\{\Mg \Hg(r^{2} M^a)\}((1+2b)\rho)|^p.
\rho^{m-1-2p}d\rho \right)^{1/p} \notag  \\
& \leq 
C (1+2b)^{2-\frac{m}{p}}\left(\int_0^\infty 
|M^a (\rho)|^p \rho^{m-1}d\rho \right)^{1/p} \notag \\
& \leq C(1+2a)^{\frac{m}p}
(1+2b)^{2-\frac{m}{p}}\|V\p\|_1 \|u\|_1. \lbeq(8-2)
\end{align}
Since $m-2-k-(j-1)\geq 2$ for $k+j\leq m-2$, it follows 
by using \refeq(kjkab-sub) that  
\begin{align} 
& \left\|
\int_{|y|>1}  
\frac{(V\f)(x-y) Q_{jk}^{a,b}(|y|)}{|y|^{m-2-k}}dy 
\right\| 
\leq C \|V\f\|_1 
\left(
\int_1^\infty 
\frac{|Q_{jk}^{a,b}(\rho)|^p} {|\rho|^{p(m-2-k)}}
\rho^{m-1}d\rho
\right)^{1/p} \notag \\
& \leq C \|V\f\|_1 
\frac{(1+2b)^{j-1}}{(1+2a)^{j+2}}
\left(
\int_1^\infty |\{\Mg \Hg(r^{2} M^a)\}((1+2b)\rho)|^p 
\rho^{m-1-2p}d\rho \right)^{1/p} \notag \\
& \leq C \|V\f\|_1 
\frac{(1+2a)^{\frac{m}{p}-j-2}}
{(1+2b)^{\frac{m}{p}-j-1}}
\|V\p\|_1 \|u\|_p.  \lbeq(ev-8)
\end{align}
For the integral over $|y|<1$, we first estimate it via
H\"older's inequality   
\begin{align}
& \int_{|y|\leq 1} 
\frac{|(V\f)(x-y)Q_{jk}^{a,b}(|y|)|}
{|y|^{m-k-2}}dy \notag \\
& \leq \left(\int_{|y|\leq 1}\frac{|(V\f)(x-y)|^{p'} dy}
{|y|^{p'(m-4)}}\right)^{1/p'}
\left(\int_{|y|\leq 1} 
\left|
\frac{Q_{jk}^{a,b}(|y|)}{|y|^{2}}
\right|^p dy \right)^{1/p}. \lbeq(8-8) 
\end{align}
We then use \refeq(kjkab-sub) for the case 
$\rho<1$ to estimate the second factor by  
\begin{align}
& C \frac{(1+2b)^{j-1}}{(1+2a)^{j+2}}
\left(
\int_0^1 |\{\Mg \Hg(r^{2} M^a)\}((1+2b)\rho)|^p 
\rho^{m-1-2p}d\rho \right)^{1/p}
\notag \\
& \hspace{2cm} \leq C \frac{(1+2b)^{2-\frac{m}{p}+j-1}}
{(1+2a)^{j+2-\frac{m}{p}}}
\|V\p\|_1 \|u\|_p.  \lbeq(8-add)
\end{align}
We apply Minkowski's inequality to \refeq(8-8). 
Since $p'\leq  p$ and $p'(m-4)<m$, \refeq(8-add) then 
implies    
\bqn 
\left\|\int_{|y|\leq 1} 
\frac{|(V\f)(x-y)Q_{jk}^{a,b}(|y|)|}
{|y|^{m-k-2}}dy \right\|_p  
\leq C\frac{(1+2a)^{\frac{m}p-j-2}}
{(1+2b)^{\frac{m}{p}-j-1}} \|V\p\|_1 \|V\f\|_p \|u\|_p . 
\lbeq(8-9) 
\eqn 
Combining \refeq(ev-8) and \refeq(8-9) and  
noticing that 
$(1+2a)^{\frac{m}p-j-2}(1+2b)^{j+1-\frac{m}{p}}$ 
is summable with respect to $T^{(a)}_j T^{(b)}_k$, 
we obtain for all $1\leq j \leq \n$ and $0\leq k \leq \n$ 
that 
\bqn 
\|Z_s^{(j,k)} u\|_p 
\leq C \|V\p\|_1 (\|V\f\|_1+ \|V\f\|_{p}) \|u\|_p. 
\lbeq(8-10) 
\eqn 
When $j=0$, in parallel with \refeq(smallr-af) we have 
\begin{align} 
Q_{0k}^{a,b}(\rho) & = \frac{i|\Si|}{(1+2a)^{2}} 
\int_{0}^\infty \lam ^{k} e^{i\lam (1+2b)\rho} 
\Fg (\widetilde{M}^a)(\lam ) 
F(\lam ) d\lam  \lbeq(k0kab-extra-1)\\
& \absleq C (1+2a)^{-2} \Mg(\widetilde{M}^a)((1+2b)\rho). 
\lbeq(k0kab-extra) 
\end{align} 
Then, change of variables, \reflm(ap) for the $A_p$ weight 
$\rho^{m-1-2p}$ for $m/3<p<m/2$ and Hardy's inequality 
imply   
\begin{align*} 
& \left(\int_{0}^\infty  
\left|\frac{Q_{0k}^{a,b}(\rho)}{\rho^{2}}\right|^p \rho^{m-1}d\rho 
\right)^{1/p} 
\leq C
\frac{(1+2b)^{2-\frac{m}{p}}}{(1+2a)^{2}}
\left( \int_0^\infty r^{m-1-2p}
|\widetilde{M^a}(r)|^p 
d\rho\right)^{1/p} \notag \\  
& \leq C 
\frac{(1+2a)^{\frac{m}{p}-2}}{(1+2b)^{\frac{m}{p}-2}}
\left( 
\int_0^\infty r^{m-1}|M(r)|^p dr
\right)^{1/p} 
\leq  
C \frac{(1+2a)^{\frac{m}{p}-2}}{(1+2b)^{\frac{m}{p}-2}}
\|V\p\|_1 \|u\|_p . 
\end{align*}
We then repeat the argument of the last part of 
the proof for the case $j \geq 1$, using that 
$m-2-k\geq 2$. This yields 
\begin{multline}
\left\|\int_{\R^m} 
\frac{|(V\f)(x-y)Q_{jk}^{a,b}(|y|)|}
{|y|^{m-k-2}}dy \right\|_p  \\
\leq C\frac{(1+2a)^{\frac{m}p-2}}
{(1+2b)^{\frac{m}{p}-2}}
(\|V\f\|_1+ \|V\f\|_{p})\|V\p\|_1 \|u\|_p . 
\end{multline}
Here $\frac{m}{p}-2-2\n<-1$. It follows that 
\[
\| Z_s^{0,k}u \|^p \leq C 
(\|V\f\|_1+ \|V\f\|_{p})\|V\p\|_1 
\|u\|_p , \quad k=0, \dots, \n. 
\]
This completes the proof of the desired 
\bqn 
\|Z_s u\|_p \leq C\|u\|_p, \quad m/3<p<m/2. 
\eqn 

\subsection{Estimate of ${Z}_{\log}$}
We still have to prove 
\bqn 
\|{Z}_{\log}u\|_p \leq C_p \|u\|_p, \quad 
1<p<m/2, \quad m \geq 6 \ \ \mbox{is even}.
\lbeq(log-sest)
\eqn 
This can be done by modifying the argument for $Z_s$ 
of the previous subsections and we explain here how to 
do it. We do it when $m=6$ as other cases are simpler. 
By virtue of \refeq(loge), it suffices to prove 
\refeq(e-et) for 
\bqn 
Z_{\log}^{\alpha,\beta}= \frac{i}{\pi}\int^\infty_0 
G_0(\lam)V \lam^{\alpha+1} (\log \lam)^{\beta} 
D_{\alpha,\beta}
(G_0(\lam)-G_0(-\lam))F(\lam) d\lam \lbeq(loge-1) 
\eqn 
when $\alpha=0,1$ and $\beta=1,2$ and $VD_{\alpha,\beta}$ 
is given by \refeq(sing-ef). The operator with the strongest 
singularity with respect to $\lam$ is the one with 
$(\alpha, \beta)=(0,2)$ and we deal with that operator 
only, omitting 
the suffix $\alpha,\beta$ from $Z_{\log}^{\alpha,\beta}$.
Inserting \refeq(sing-ef) for $VD_{0,2}$ will produce 
$2d$ number of operators and we prove any of them 
satisfies \refeq(log-sest), which we still denote by  
\bqn 
Z_{\log}u= \frac{i}{\pi}\int^\infty_0 
G_0(\lam)|\f \ra \la \p | 
(G_0(\lam)-G_0(-\lam))\lam (\log \lam)^{2} F(\lam) d\lam. 
\lbeq(loge-2) 
\eqn 
It is important to recall that $\f, \p$ satisfy 
$\f,\p \in L^1 \cap L^6(\R^6)$. 
The argument at the beginning of Section 4 shows 
that $Z_{\log}$ is the sum of $Z_{\log}^{(j,k)}$, 
$0\leq j,k\leq \n$ which are defined by the second 
member of \refeq(Wsm-rep) with following changes: 
\ben 
\item[{\rm (1)}] Change $V\f$ in \refeq(Wsm-rep) 
and $V\p$ inside $M(r)$ by $\f$ and $\p$ of 
\refeq(sing-ef) respectively.
\item[{\rm (2)}] Change $\lam^{j+k-1}$ or $\lam^{k}$ 
in the definition \refeq(kjkab) or \refeq(k0kab) of 
$Q_{jk}^{a,b}(\rho)$ by 
$\lam^{j+k+1}(\log \lam)^{2}$ or 
$\lam^{k+2}(\log \lam)^{2}$ respectively. 
\een
Then, we can check the following:   
\ben 
\item[{\rm (i)}]  \reflm(zsq2) holds with 
$Z_{\log}^{(j,k)}$ in place of $Z_{s}^{(j,k)}$ for all 
$0\leq j,k \leq \n$ such that $(j,k)\not=(\n,\n)$, 
hence, a fortiori, so does \reflm(zsq3).    
This is because {\rm (a)} the conjugate Fourier transforms of 
derivatives of order up to $k+1$ of 
$\lam^{j+k+1} (\log \lam)^2 F(\lam)$ for $j\geq 1$ 
or of $\lam^{k+2} (\log \lam)^2 F(\lam)$ which 
will appear in the integration by parts formula
corresponding to \refeq(devel) for $j\geq 1$ or 
\refeq(fg-1) have symmetric decreasing integrable 
majorants, which is sufficient for obtaining \refeq(devel-a) 
or \refeq(fg-1) and because {\rm (b)} 
$\f,\p \in L^1 \cap L^6(\R^6)$, which is sufficient to 
obtain \reflm(njl) and, e.g. \refeq(d6-0) and \refeq(d7-m). 
\item[{\rm (2)}] \refprp(last-prop) holds with 
$Z_{\log}^{(\n,\n)}$ in place of $Z_{s}^{(\n,\n)}$. 
To see this, we first remark that in the 
equation corresponding to \refeq(new) the conjugate Fourier 
transforms of the derivatives 
of $\lam^{2\n+1}(\log\lam)^2 F(\lam)$ upto the order 
$\n$ have symmetric decreasing integrable majorants. 
This and $\f,\p \in L^1 \cap L^6(\R^6)$ implies 
\refeq(Ll3) for the operators corresponding to 
$Z_{\log}$. We then need study the operator 
\refeq(m-1) with $\lam^{m-1} (\log\lam)^2$ in place 
of $\lam^{m-3}$ (recall we are assuming $m=6$). 
By this change $\lam^{m-4}$ in \refeq(f-1) is replaced by 
$\lam^{m-2} (\log\lam)^2$. If we change 
$\lam^\sigma$ by $\lam^\sigma (\log)^2$ in the definition 
\refeq(dlt) of $\tilde{L}_\pm (\rho, r)$, then the 
conclusion \refeq(dlt-con) holds with 
$\la \rho-r\ra^{-\sigma}$ in place of 
$\la \rho-r\ra^{-(\sigma+1)}$. 
Thus, we have in the estimates corresponding 
\refeq(g-13) and \refeq(g-13a) the faster decaying factor 
$\la \rho-r\ra^{-(m-2)}$ in place of 
$\la \rho-r\ra^{-(m-3)}$.
This produces \refeq(gfnal) and we obtain 
\refprp(last-prop) for $Z_{\log}^{(\n,\n)}$.
\item[{\rm (3)}] With the modification as in (1) above, 
\refeq(log-sest) for $m/3<p<m/2$ may be proved 
by repeating the argument of Section 4.2 with almost 
no changes.  
\een



\begin{thebibliography}{99}

\bibitem{Ag} S. Agmon, 
\textit{Spectral properties of Schr\"odinger
operators and scattering theory}, Ann. Scuola 
Norm. Sup. Pisa Cl. Sci. (4)
 {\bf 2} (1975), 151--218. 

\bibitem{AS} M. Aizenman and B. Simon, 
{\it Brownian motion and Harnack inequality 
for Schr\"odinger equations}, 
Comm. Pure Appl. Math. {\bf 35} (1982), 209–273.


\bibitem{AY} G. Artbazar and K. Yajima, 
\textit{The $L^p$-continuity of 
wave operators for one dimensional Schr\"odinger 
operators}, J. Math. Sci. 
Univ. Tokyo {\bf 7} (2000), 221-240. 


\bibitem{Be} M. Beceanu, {\it Structure of wave operators 
for a scaling-critical class of potentials}, 
Amer. J. Math. {\bf 136} (2014), 255--308. 

\bibitem{BL} J. Bergh and J. L\"ofstr\"om, 
\textit{Interpolation spaces, an introduction}, 
Springer Verlag, 
Berlin-Heidelberg-New York (1976). 

\bibitem{DF} P. D'Ancona and L. Fanelli, 
{\it $L^p$--boundedness of the wave operator for the 
one dimensional Schr\"odinger operator}, 
Commun. Math. Phys. {\bf 268} (2006), 415--438. 

\bibitem{FY} D. Finco and K. Yajima, {\it The 
$L^p$ boundedness of wave operators for Schr\"odinger 
operators with threshold singularities. II. 
Even dimensional case}. J. Math. Sci. Univ. Tokyo 
{\bf 13} (2006), 277–346. 

\bibitem{Gr} L. Grafakos, 
{\it Modern Fourier analysis}, Springer Verlag, 
New York (2009). 


\bibitem{Jensen} A. Jensen, 
{\it Spectral properties of Schr\"odinger operators and 
time decay of the wave functions, Results in $L_2(\R^m)$, 
$m\geq 5$}, Duke Math. J. {\bf 47} (1980), 57--80. 

\bibitem{JK} A. Jensen and T. Kato, 
{\it Spectral properties of Schr\"odinger operators 
and time-decay of the wave functions}, 
Duke Math. J. {\bf 46} (1979), 583--611. 

\bibitem{JY} A. Jensen and K. Yajima 
{\it A remark on $L^2$-boundedness of wave operators 
for two dimensional Schr\"odinger operators}, Commun. 
Math. PHys. {\bf 225} (2002), 633--637. 


\bibitem{Kato-e} T. Kato, {\it Growth 
properties of solutions 
of the reduced wave equation with a 
variable coefficient}, Comm. Pure 
Appl. Math {\bf 12} (1959), 403--425.  


\bibitem{kb} T. Kato, {\it Perturbation theory 
for linear operators}, 
Springer Verlag, New York (1966). 

\bibitem{KatoS1} T. Kato, {\it Wave 
operators and similarity for 
non-selfadjoint operators}, 
Ann. Math. {\bf 162} (1966), 258--279. 


\bibitem{Ku-0} S. T. Kuroda, {\it Introduction 
to Scattering Theory}, 
Lecture Notes, Aarhus University

\bibitem{Mu} M. Murata, \textit{Asymptotic 
expansions in time for solutions
of Schr\"{o}dinger-type equations}, J. Funct. Anal., 
{\bf 49} (1982), 10--56. 


\bibitem{RSi} M. Reed and B. Simon, 
{\it Methods of moderm mathemtical physics vol II, 
Fourier analysis, selfadjointness}, Academic Press, 
New-York, San Francisco, London (1975). 

\bibitem{RS3} M. Reed and B. Simon, 
{\it Methods of moderm mathemtical physics vol III, 
Scattering theory}, Academic Press, New-York, 
San Francisco, London (1979). 

\bibitem{RS4} M. Reed and B. Simon, 
{\it Methods of moderm mathemtical physics vol IV, 
Analysis of Operators}, Academic Press, New-York, 
San Francisco, London (1978). 



\bibitem{Stempak}, K. Stempak{it Transplantation 
theorems—a survey}, J. Fourier Anal. Appl. {\bf 17} 
(2011), 408–430. 


\bibitem{Stein} E. M. Stein, {\it Harmonic analysis, 
real-variable methods, orthogonality, and 
oscillatory integrals}, Princeton Univ. Press, 
Princeton, NJ. (1993). 

\bibitem{Weder} R. Weder, 
\textit{$L^p$-$L^{p'}$ estimates for the 
Schr\"odinger equations on the line and 
inverse scattering for the nonliner 
Schr\"odinger equation with a potential}, 
J. Funct. Anal. {\bf 170} 
(2000), 37--68. 

\bibitem{WW} E. T. Whittaker and G. N. Watson, 
{\it A Course of Modern Analysis}. 
Cambridge University Press; 4th edition (1927). 

\bibitem{Y-d3} K. Yajima,  \textit{The 
$W^{k,p}$-continuity of wave operators for 
Schr\"odinger operators}, J. Math. Soc. 
Japan {\bf 47} (1995), 551-581.

\bibitem{Y-d4} K. Yajima,  
\textit{The $W^{k,p}$-continuity of wave operators 
for Schr\"odinger operators III}, J. Math. Sci. 
Univ. Tokyo {\bf 2} (1995), 311--346. 

\bibitem{Y-d2} K. Yajima, 
\textit{The $L^{p}$-boundedness of wave operators 
for two dimensional Schr\"odinger operators}, 
Commun. Math. Phys. {\bf 208} (1999), 125--152. 



\bibitem{Y-odd} K. Yajima, {\it The $L^p$ boundedness of wave 
operators for Schr\"odinger operators with threshold 
singularities I, Odd dimensional case}, J. Math. Sci. 
Univ. Tokyo {\bf 7} (2006), 43--93. 

\end{thebibliography}
\end{document}